\documentclass[prd,amssymb,amsmath,nofootinbib]{revtex4}

\begin{document}

\title{Stability of de Sitter spacetime under isotropic perturbations
in semiclassical gravity}
\author{Guillem P\'erez-Nadal}
\affiliation{Departament de F\'{\i}sica Fonamental, Universitat
de Barcelona, Av.~Diagonal 647, 08028 Barcelona, Spain}
\author{Albert Roura}
\affiliation{Theoretical Division, T-8, Los Alamos National Laboratory,
M.S.~B285, Los Alamos, NM 87545}
\author{Enric Verdaguer}
\affiliation{Departament de F\'{\i}sica Fonamental
and Institut de Ci\`encies del Cosmos, Universitat de Barcelona,
Av.~Diagonal 647, 08028 Barcelona, Spain}

\begin{abstract}
  A spatially flat Robertson-Walker spacetime driven by a cosmological
  constant is non-conformally coupled to a massless scalar field. The
  equations of semiclassical gravity are explicitly solved for this
  case, and a self-consistent de Sitter solution associated with the
  Bunch-Davies vacuum state is found (the effect of the quantum field
  is to shift slightly the effective cosmological constant).
  Furthermore, it is shown that the corrected de Sitter spacetime is
  stable under spatially-isotropic perturbations of the metric and the
  quantum state.  These results are independent of the free
  renormalization parameters.
\end{abstract}


\maketitle

\section{Introduction}

\label{sec1}

Our present understanding of cosmology assumes that the universe
underwent a short period of accelerated expansion known as inflation
\cite{guth81,linde82a,albrecht82,linde83a,linde90}.  The inflationary
scenario has been remarkably successful in explaining the observed
anisotropies of the cosmic microwave background
\cite{smoot92,bennett03,peiris03,spergel07}. In most inflationary
models the accelerated expansion phase is close to but never exactly
de Sitter and this phase eventually ends when the kinetic energy of
the inflaton field driving inflation starts to dominate over the
potential term. On the other hand, observations of distant supernovae
indicate that the universe is presently undergoing a period of
accelerated expansion \cite{perlmutter99,riess98} that may be driven
by a small non-vanishing cosmological constant
\cite{seljak06,tegmark06,giannantonio06,eisenstein05}. If that is the
case, the geometry of our universe would tend to that of de Sitter
spacetime at sufficiently late times. Thus, a detailed knowledge of
the physics associated with de Sitter space may play a key role in
understanding both the very early universe as well as its ultimate
fate. Furthermore, it is conceivable that studying a possible
screening of the cosmological constant driving de Sitter space, due to
quantum effects, could shed some light on the huge fine-tuning problem
that the current value of the cosmological constant seems to pose.

An open question which has recently received increasing attention is
whether the quantum fluctuations of the metric and the matter fields
in de Sitter space can give rise to large back-reaction effects on the
mean background geometry. It has been argued that in pure gravity with
a cosmological constant the infrared effects due to two graviton loops
and higher-order radiative corrections could lead to a secular
screening of the cosmological constant
\cite{tsamis96a,tsamis97}. There have also been proposals that a
significant screening of the cosmological constant could appear in
chaotic inflationary models at one loop when both the metric and
inflaton field fluctuations are considered
\cite{mukhanov97,abramo97,abramo99,losic05}. In all these cases the
quantum fluctuations of the metric play an essential role. However,
whenever the metric perturbations are quantized, one needs to confront
the problem of defining proper diffeomorphism-invariant observables in
quantum gravity \cite{giddings05}, even when treated as a low-energy
effective field theory. In particular one needs to make sure that the
secular screening found in the analysis mentioned above is not simply
a gauge artifact. As a matter of fact, it was shown in
Refs.~\cite{abramo02b,geshnizjani02} that when a suitable
gauge-invariant measure of the expansion rate was considered the
screening effect previously found in chaotic inflationary models was
not actually present (at least for single field models). Similarly, a
recent reanalysis of the pure gravity case which made use of a
diffeomorphism-invariant measure of the change of the expansion rate
revealed the absence of secular effects to all orders in perturbation
theory \cite{garriga07}.

In recent work it was also found that the back reaction due to
one-loop effects of massless non-conformal fields can give rise to
substantial deviations from de Sitter spacetime
\cite{espriu05,cabrer07}. This would be the case even if the quantum
fluctuations of the metric are not considered (provided that there is
some other massless non-conformal field in addition to gravitons). Not
quantizing the metric perturbations means that the ambiguity
associated with the gauge-fixing term for the metric perturbations is
no longer present, and the mean geometry is a perfectly well-defined
(and gauge-independent) object. Hence, the subtleties mentioned in the
previous paragraph do not apply.  The heuristic argument provided in
Refs.~\cite{espriu05,cabrer07} is that one-loop contributions from
massless non-conformal fields correspond to logarithmic non-local
terms (conformal fields only produce local terms for Robertson-Walker
geometries), and by analogy with the situation in pion physics, one
expects that they become important in the infrared limit.  However, it
is not clear that pion physics constitutes a good analogy because the
derivative coupling of the matter fields to the metric generates
higher powers of the momentum. This point can be illustrated with the
simple example of small metric perturbations around flat space. In
that case the Fourier transform of the inverse propagator behaves like
$k^2 [1 + b\, (k^2 / m_\mathrm{p}^2) \ln (k^2/\mu_0^2)]$, where $b$ is
a dimensionless number roughly of order one, $m_\mathrm{p}$ is the
Planck mass and $\mu_0$ is some fixed mass scale. One can see that
although the logarithm grows in the infrared, the whole term actually
decreases because it is suppressed by the factor $(k^2 /
m_\mathrm{p}^2)$. Indeed, a detailed calculation of the quantum
radiative corrections to the Newtonian potential shows that the
contribution from that term is suppressed by the square of the ratio
of the Planck length over the radial distance
\cite{donoghue94a,donoghue94b,bjerrum03}. Of course the case of a
Robertson-Walker (RW) metric, which involves a time-dependent scale
factor, is not so simple and deserves a careful analysis in order to
compare with the detailed calculation in
Refs.~\cite{espriu05,cabrer07}.

In this paper we approach this problem by explicitly solving the back
reaction on the mean gravitational field due to the quantum effects of
a massless non-conformally coupled scalar field when the quantum
fluctuations of the metric are not considered. This kind of one-loop
calculation is entirely equivalent to studying the corresponding
back-reaction problem in the semiclassical gravity framework
\cite{birrell94,wald94,flanagan96} by solving self-consistently the
semiclassical Einstein equation, which includes the suitably
renormalized quantum expectation value of the stress tensor operator
acting as a source.  Specifically, in our calculation we assume the
presence of a cosmological constant, which would lead to a de Sitter
solution in the absence of quantum effects, and simplify the problem
by focusing on RW geometries, corresponding to spatially homogenous
and isotropic states of the quantum field.

There exist relevant antecedents to our analysis in the context of
quantum field theory in a fixed curved spacetime, \emph{i.e.}, when
the back reaction of the quantum fields on the spacetime geometry is
not taken into account. The so-called Bunch-Davies vacuum
\cite{bunch78a,birrell94} for fields in de Sitter is a state invariant
under all the isometries of de Sitter space, which is maximally
symmetric. The renormalized expectation value of the stress tensor
operator for that state is proportional (with a constant factor) to
the metric and therefore its contribution to the semiclassical
Einstein equation has the same form as a cosmological constant
term. More importantly, it was shown in Ref.~\cite{anderson00} that
for fields with a wide range of mass and curvature-coupling parameters
evolving in a given de Sitter spacetime, the expectation value of the
stress tensor for any reasonable initial state
tended at late times to the expectation value for the Bunch-Davies
vacuum, where by reasonable states one means states with the same
ultraviolet behavior as the Minkowski vacuum, \emph{i.e.}, with
essentially no excitations at arbitrarily high frequencies (technically they are known as Hadamard states \cite{birrell94,wald94}). This
result can be intuitively understood as follows: the exponential
expansion will redshift any finite frequency excitations of the
Bunch-Davies vacuum so that their contribution to the stress tensor
will tend to zero at late times.

The result described in the previous paragraph suggests that even when
taking into account back-reaction effects, perturbations around de
Sitter will be redshifted away and at late times the spacetime
geometry will approach de Sitter space with an effective cosmological
constant which includes the contribution from the expectation value of
the stress tensor for the Bunch-Davies vacuum. However, in order to
prove this expectation without any room for doubt, one needs to solve
both the semiclassical Einstein equation and Klein-Gordon equation for
the scalar field self-consistently. This is the main goal of this
paper. We will consider a fairly general family of Gaussian initial
states for the quantum field which are spatially homogenous and
isotropic, and discuss under what conditions the trace of the
stress-tensor expectation value exhibits unphysical divergences at the
initial time after the standard renormalization procedure. We will
also explain how to select appropriate states with regular initial
behavior, and then solve the back-reaction equation
explicitly. The standard renormalization procedure for the ultraviolet
divergencies of the expectation value of the stress tensor requires
the renormalization of the gravitational and cosmological constants as
well as two new dimensionless parameters which are related to local
geometric terms in the gravitational action which are quadratic in the
curvature tensor. These new parameters should in principle be
determined experimentally in order to eliminate the two-parameter
ambiguity otherwise exhibited by the back-reaction
equation. Nevertheless, for the particular case that we are
considering (and at the order in the Planck length at which we are
working) the results turn out to be independent of the particular
value of these renormalization parameters.

The plan of the paper is the following. In Sec.~\ref{sec2} we
introduce the particular model that we will be considering and
describe the procedure that we will employ to generate the initial
state of the quantum field. In Sec.~\ref{sec3} we use the Closed Time
Path (CTP) formalism, whose application to semiclassical gravity is
briefly summarized in the Appendix, 
to derive the effective action which encodes the effect of the quantum
fields on a spatially flat RW spacetime. We also derive the relation
between the time-time component of the stress tensor and its trace,
which will be later used to derive the back-reaction equation. In
Sec.~\ref{sec4} the CTP effective action is used to obtain the quantum
expectation value of the stress tensor operator, and the suitable
initial conditions for the quantum state of the field and for the
cosmological scale factor are discussed in detail. Finally, in
Sec.~\ref{sec5} the semiclassical Friedmann equation that describes
the back reaction of the quantum field on the scale factor, driven by
a cosmological constant, is derived and solved perturbatively in
powers of the Planck length over the Hubble radius. We find an
explicit result valid for all times. It shows that there exists a
self-consistent de Sitter solution with a slightly shifted
cosmological constant due to the one-loop effects and that all the
other solutions tend to this one at late times.

\section{A quantum field in a RW background and its initial state}
\label{sec2}

In this section we describe our model for the back reaction of quantum
fields on a cosmological background.  We assume a spatially
homogeneous and isotropic cosmological model with flat spatial
sections, described by the metric
\begin{equation}
g_{\mu\nu}=a^2(\eta)\eta_{\mu\nu},
\label{2.1}
\end{equation}
where $\eta_{\mu\nu}$ is the $n$-dimensional Minkowski metric (we use
arbitrary dimensions for the moment in order to perform dimensional
regularization later on), which takes the form
$\eta_{\mu\nu}=\rm{diag}(-1, 1,\dots,1)$ when considering the usual
inertial coordinates, and $a(\eta)$ is the cosmological scale factor
in terms of the conformal time $\eta$, which is related to the
physical time $t$ by $a \, d\eta=dt$. The classical action for a real
massless scalar field $\Phi(x)$ coupled to gravity is
\begin{equation}
S_\mathrm{m}[g_{\mu\nu},\Phi]=-\frac{1}{2}\int d^n x\sqrt{-g}\left[
g^{\mu\nu}\partial_\mu\Phi\partial_\nu\Phi +(\xi_c+\nu)R \, \Phi^2\right],
\label{2.2}
\end{equation}
where the dimensionless parameters $\xi_\mathrm{c}\equiv
(n-2)/[4(n-1)]$ (equal to $1/6$ in four dimensions) and $\nu$ give the
coupling to the Ricci curvature scalar $R$, given in this case by
\begin{equation}
R=2(n-1)\left(\frac{\ddot a}{a^3}+\frac{n-4}{2}\frac{\dot a^2}{a^4}\right).
\label{2.3}
\end{equation}
where here and throughout the rest of the paper overdots denote
derivatives with respect to the conformal time, \emph{i.e.}, $\dot{}
\equiv d/d\eta$.  The \emph{minimal coupling} case (no direct coupling
to the curvature) corresponds to $\nu=-\xi_\mathrm{c}$; a massless
scalar field with minimal coupling mimics the behavior of gravitons in
the cosmological background, except for a factor of two corresponding
to the graviton polarizations. When $\nu=0$ the classical action
$S_\mathrm{m}[g_{\mu\nu},\Phi]$ is invariant under conformal
transformations with $g_{\mu\nu} \to \Omega^2(x) \, g_{\mu\nu}$ and
$\Phi(x) \to \Omega^{(2-n)/2}(x) \, \Phi(x)$; this is known as the
\emph{conformal coupling} case and it can be used to mimic the
behavior of photons.

Since the RW metric given by Eq.~(\ref{2.1}) is conformally flat, it
is convenient to introduce the rescaled scalar field
$\phi(x)=a^{(n-2)/2}(\eta) \, \Phi(x)$. The action in Eq.~(\ref{2.2})
then simplifies to
\begin{equation}
S_\mathrm{m}[a,\phi]=-\frac{1}{2}\int d^n x
\left(\eta^{\mu\nu}\partial_\mu\phi\partial_\nu\phi+\nu a^2R \, \phi^2\right),
\label{2.4}
\end{equation}
which is the action for a free scalar field $\phi(x)$ in Minkowski
spacetime with a time-dependent quadratic coupling $\nu a^2 R \,
\phi^2$. Identifying the matter Lagrangian $L_\mathrm{m}$ from
$S_\mathrm{m}=\int d\eta L_\mathrm{m}$ and the momentum $\pi(x)=\delta
L_\mathrm{m} / \delta \dot\phi (x) = \dot\phi (x)$, one obtains the
following Hamiltonian for the rescaled scalar field:
\begin{equation}
H_\mathrm{m}[\pi,\phi]=\frac{1}{2}\int d^{n-1} x
\left[\pi^2+\left(\vec\nabla\phi\right)^2+\nu a^2R \, \phi^2\right]
\equiv H_\mathrm{m}^{(0)}+H_{\textrm{int}},
\label{2.5}
\end{equation}
where in the second equality we have separated the Hamiltonian into
that of a free massless field, $H_\mathrm{m}^{(0)}$, and an
interaction Hamiltonian $H_{\textrm{int}}$, which is proportional to
$\nu$.

Let us now discuss the kind of initial quantum states of the field
that we will be considering.
We are interested in the evolution of the scale factor driven by a
cosmological constant $\Lambda$ plus the back-reaction effect due to
the quantum scalar field, given some initial conditions for the scale
factor and its derivative at some initial time $\eta_\mathrm{i}$ as
well as the initial state of the quantum field at that time. On the
other hand, we will use the evolution from $\eta=-\infty$ to
$\eta=\eta_\mathrm{i}$ as an auxiliary way to prepare the initial
quantum state of the field. More precisely, the initial state of the
field will be given by
\begin{equation}
|\Psi_\mathrm{i}\rangle=\hat{U}(\eta_\mathrm{i},-\infty) \, |0,-\infty\rangle,
\label{2.15}
\end{equation}
where $|0,-\infty\rangle$ is the usual Minkowski vacuum at
$\eta\to-\infty$ and $\hat{U}$ is the time evolution operator
associated with the Hamiltonian given by
Eq.~(\ref{2.5}). $\hat{U}(\eta_\mathrm{i},-\infty)$ only depends on
the scale factor before the initial time $\eta_\mathrm{i}$, which will
be denoted from now on by $a_\Psi(\eta)$ and can be a fairly arbitrary
regular function, subject only to condition
\begin{equation}
\lim_{\eta\to -\infty} a_\Psi^2(\eta) \, R_\Psi(\eta) = 0,
\label{2.6}
\end{equation}
where $R_\Psi(\eta)$ is the Ricci scalar corresponding to
$a_\Psi(\eta)$, and to the requirement of a sufficiently smooth
transition at $\eta_\mathrm{i}$ to the scale factor at later times
(the reason for this latter condition will be explained in detail in
Sec.~\ref{sec4}).  Thus, our initial state is a squeezed state%
\footnote{This means that we restrict our attention to a family of
  Gaussian pure states.}  that evolves (in the Schr\"odinger picture)
from the Minkowski vacuum state. A particular case is the Bunch-Davies
vacuum \cite{bunch78a} for zero mass and curvature coupling
$(\xi_\mathrm{c} + \nu)$, which would follow from considering a scale
factor $a_\Psi(\eta)$ that corresponds to a given de Sitter spacetime
all the way from $\eta=-\infty$ to $\eta=\eta_\mathrm{i}$, [note that
such a scale factor does satisfy condition (\ref{2.6})].  Note that
the state obtained from the construction described above and defined
by Eq.~(\ref{2.15}) is the state of the rescaled field
$\phi(x)$. However, the state of the original field $\Phi(x)$ can be
derived straightforwardly from it if one takes into account the simple
relation between $\phi(x)$ and $\Phi(x)$ involving the scale factor
$a(\eta)$.

Finally, the gravitational action for the scale factor can be written
as
\begin{equation}
S_\mathrm{g}[a]=\frac{\cal V}{2\kappa}\int_{\eta_\mathrm{i}}d\eta \,
a^n (R-2\Lambda) +S_\mathrm{g}^\mathrm{c}[a],
\label{2.7}
\end{equation}
where the first term is just the Einstein-Hilbert term with
$\kappa=8\pi G=8\pi/m_\mathrm{p}^2$ ($G$ is the gravitational coupling
constant, $m_\mathrm{p}$ is the Planck mass and we are using natural
units with $\hbar=c=1$), ${\cal V}=\int d^{n-1}x$ is a spatial
comoving volume factor that will drop in the final expressions, and
$S_\mathrm{g}^\mathrm{c}[a]$ accounts for the gravitational
counterterms that will be specified later. At this point the
parameters $\kappa$ and $\Lambda$ should in principle be considered as
bare parameters. However, for a massless field these parameters do not
need to be renormalized when using dimensional regularization since
the divergencies in that case only require counterterms which are
quadratic in the curvature, as we will see below.


\section{The effective action for the cosmological scale factor}
\label{sec3}

\subsection{The expectation value of the energy density and
the trace of the stress tensor}
\label{sec3.1}

We can now follow the procedure outlined in the appendix 
to derive the semiclassical Einstein equation describing the back
reaction of the scalar field on the spacetime geometry. The
expectation value of the stress tensor for a given state of the field
plays a key role in that equation. As long as one considers spatially
homogenous and isotropic states of the quantum fields, it is
consistent to assume that the metric in the semiclassical equation
(\ref{2.8}) takes the restricted form (\ref{2.1}) throughout, so that
there is only one dynamical variable $a(\eta)$ to be determined.

Hence, we can concentrate on just one of the equations for the
different components of Eq.~(\ref{2.8}), and in particular on the $00$
component, which in this semiclassical cosmological context may be
called the \emph{semiclassical Friedmann equation}. The expectation
value $\langle \hat{T}_{00}\rangle_\mathrm{ren}$ will be taken in the
state $|\Psi_\mathrm{i}\rangle$ defined by Eq.~(\ref{2.15}).  Both the
classical stress tensor and its quantum expectation value can be
obtained by functionally differentiating, respectively, the classical
action $S_\mathrm{m}$ and the influence action $S_{\mathrm{IF}}$ with
respect to the metric, according to Eqs.~(\ref{2.12}) and
(\ref{2.12a}). The influence action describes the effect of the
quantum matter fields on the gravitational field and results from
functionally integrating the quantum matter fields.  Note, however,
that since our metric has been assumed to have the from
$g_{\mu\nu}=a^2(\eta)\eta_{\mu\nu}$, with the scale factor as the only
independent kinematical degree of freedom, we can only functionally
differentiate with respect to $a(\eta)$ and will just be able to
obtain trace of the stress tensor, $T^\mu_\mu$. This can be seen as
follows. Since $\delta g_{\mu\nu}=2 a\eta_{\mu\nu}\delta
a=2a^{-1}g_{\mu\nu}\delta a$, from Eq.~(\ref{2.12}) we can write
\begin{equation}
\delta S_\mathrm{m}\equiv\int d^nx\frac{\delta S_\mathrm{m}}{\delta g_{\mu\nu}}\delta g_{\mu\nu}
=\int d^nx\sqrt{-g} \, T^{\mu\nu}g_{\mu\nu}\frac{\delta a}{a}=\int d\eta \int d^{n-1}x \,
a^{n-1}T^\mu_\mu \, \delta a,
\label{2.16}
\end{equation}
so that $\delta S_\mathrm{m}/\delta a={\cal V}a^{n-1}T^\mu_\mu$, where
the spatial comoving volume ${\cal V}$ appears as a consequence of the
spatial homogeneity of the stress tensor. This also means that by
functional derivation with respect to the scale factor we obtain, from
the renormalized influence action $S^{\mathrm{ren}}_{\mathrm{IF}}$ in
Eq.~(\ref{2.12a}), the renormalized expectation value of the trace of
the stress tensor operator:
\begin {equation}
\langle \hat{T}^\mu_\mu\rangle=  \frac{1}{{\cal V}a^{n-1}}
\left.\frac{\delta S^{\mathrm{ren}}_{\textrm{IF}}[a^+,a^-]}{{\delta a}^+}\right|_{a^+=a^-=a},
\label{2.17}
\end{equation}
where from now on we will drop the subscript ``ren'' in the
expectation value to simplify the notation.

In principle one could use either the $00$ component of the
semiclassical Einstein equation (the Friedmann equation) or the
equation for the trace. There is, however, a subtle difference. For
the sake of the argument let us consider the classical limit only;
whereas the classical Friedmann equation is a first-order differential
equation in time for the scale factor the equation for the trace is of
second order in time. The solutions of the Friedmann equation
automatically satisfy the trace equation, but the Friedmann equation
also constrains the initial conditions for the scale factor and its
time derivative. If one works only with the trace equation, this
additional information is missed. (When quantum corrections are added,
higher-order derivatives appear in both equations, but we will explain
how to deal with them in Sec.~\ref{sec5}.)

In this paper we will work with the semiclassical Friedmann
equation. Therefore, we need to calculate $\langle
\hat{T}_{00}\rangle$. One possibility is to start with a metric of the
form $ds^2 = - N^2(\tilde{\eta}) d\tilde{\eta}^2 + a^2(\tilde{\eta})
\delta_{ij} dx^i dx^j$ with $N(\tilde{\eta})$ and $a(\tilde{\eta})$
independent,%
\footnote{Provided that $N(\tilde{\eta})$ is non-vanishing and
  differentiable, such a metric can always be rewritten as $ds^2 =
  a^2(\eta) ( - d\eta^2 + \delta_{ij} dx^i dx^j)$ through a coordinate
  transformation involving a redefinition of the time $\tilde{\eta}$.}
functionally differentiate with respect to $N$ and $a$, and finally
take $N=a$ only after that. The functional derivative with respect to
$N$ gives $\langle \hat{T}_{00}\rangle$ and the Friedmann equation.
Alternatively, one can make use of a useful relation between $\langle
\hat{T}_{00}\rangle$ and $\langle \hat{T}^\mu_\mu\rangle$ in a RW
spacetime which is a consequence of the stress-tensor conservation law
$\nabla_\mu \langle \hat{T}^{\mu\nu}\rangle=0$, and the fact that
$\vec \xi=\partial/\partial\eta$ is a conformal Killing field,
\emph{i.e.}, $2\nabla_{(a}\xi_{b)}=\lambda g_{\mu\nu}$ with
$\lambda=2\dot a/a$ in our case \cite{parker79,birrell94}. These two
equations lead to
\begin{equation}
\nabla_\mu\left(\langle \hat{T}^{\mu\nu}\rangle\xi_\nu\right)=\frac{\dot a}{a}\langle \hat{T}^\mu_\mu\rangle,
\label{2.18}
\end{equation}
which can be integrated over the spacetime volume bounded by the
space-like hypersurfaces corresponding to $\eta'=-\infty$ and
$\eta'=\eta$ (for the construction introduced in Sec.~\ref{sec2} to
generate the initial quantum state of the field the stress tensor is
also conserved from $\eta'=-\infty$ till our initial time
$\eta'=\eta_\mathrm{i}$). Using Gauss's theorem we get a relationship
between the integration of $\langle \hat{T}_{00}\rangle$ on the two
hypersurfaces and the spacetime integral of $\langle
\hat{T}^\mu_\mu\rangle$ which reads
\begin{equation}
\langle \hat{T}_{00}(\eta)\rangle a^{n-2}(\eta)=C-\int^{\eta}_{-\infty}d\eta^\prime  a^{n-1}(\eta^\prime)\dot a(\eta^\prime) \langle \hat{T}^\mu_\mu(\eta^\prime)\rangle,
\label{2.19}
\end{equation}
where $C=\langle \hat{T}_{00}(-\infty)\rangle a^{n-2}(-\infty)$ and we
have divided by the spatial volume ${\cal V}$, which appears due to
the spatial homogeneity of the stress-tensor expectation value. Since
the constant $C$ is proportional to the expectation value of the
energy density of a Minkowski vacuum, as follows from
Eq.~(\ref{2.15}), it should vanish. Hence, from now on, we will take
$C=0$.

\subsection{The effective action}
\label{sec3.2}

In this subsection we compute the influence action needed to derive
the expectation value of the trace of the stress tensor according to
Eq.~(\ref{2.17}). Using the so-called closed time path (CTP)
formalism, the influence action $S_{\textrm{IF}}[g^+,g^-]$ for an
arbitrary metric is defined in the appendix 
by Eqs.~(\ref{2.10}) or (\ref{2.13}) for a general initial state of
the field. Specializing Eq.~(\ref{2.13}) to the conformally flat
metric $g_{\mu\nu} = a^2 \eta_{\mu\nu}$ with the definition of our
initial state as given in Eq.~(\ref{2.15}), we get
\begin{equation}
e^{iS_{\textrm{IF}}[a^+,a^-]}=\int {\cal D}\varphi \, \langle 0, -\infty |\hat{U}_-(-\infty,\eta_\mathrm{f})|\varphi \rangle \langle \varphi |\hat{U}_+(\eta_\mathrm{f},-\infty)|0, -\infty \rangle,
\label{2.19a}
\end{equation}
where
$|\varphi\rangle$ are the properly normalized field eigenstates, such
that $\hat\varphi (\vec{x})|\varphi(\vec{x}')\rangle=\varphi(\vec{x})
|\varphi(\vec{x}')\rangle$, and the time evolution operator is
\begin{equation}
\hat{U}_\pm(\eta_\mathrm{f},-\infty)=T\exp\left(
-i\int_{-\infty}^{\eta_\mathrm{f}}d\eta
\hat{H}_\mathrm{m}[a_\pm,\hat{\pi},\hat{\phi}]\right).
\label{2.20}
\end{equation}
Using the path integral representation for the time evolution operator
in terms of the action (\ref{2.4}) for the scalar field we have
\begin{equation}
e^{iS_{\textrm{IF}}[a^+,a^-]}= \int {\cal D}\phi^+ {\cal D}\phi^-
e^{i\left(S_\mathrm{m}[a^+,\phi^+]- S^*_\mathrm{m}[a^-,\phi^-]\right)},
\label{2.21}
\end{equation}
where $\phi^+(\eta_\mathrm{f})=\phi^-(\eta_\mathrm{f})$. This
expression corresponds to Eq.~(\ref{2.10}) for an initial Minkowski
vacuum state. To enforce this state we must take the usual
$-i\epsilon$ prescription. Integrating by parts and taking into
account that the action $S_\mathrm{m}$ is quadratic in the field, we
can write
\begin{equation}
e^{iS_{\textrm{IF}}[a^+,a^-]}= \int {\cal D}\phi^+ {\cal D}\phi^-
e^{\frac{i}{2}\int d^nx\left(\phi^+ A_{++}\phi^+
+\phi^- A_{--}\phi^-\right)}=(\det A)^{-1/2},
\label{2.22}
\end{equation}
where the matrix $A$ is defined by
$A_{++}=\eta^{\mu\nu}\partial_\mu\partial_\nu-\nu (a^+)^2R^+
+i\epsilon$, $A_{--}=-\left(\eta^{\mu\nu}\partial_\mu\partial_\nu-\nu
  (a^-)^2R^- -i\epsilon\right)$, and $A_{+-}=A_{-+}=0$, and a Gaussian
integration has been performed in the last equality. Introducing the
inverse matrix $G=A^{-1}$ we thus have
\begin{equation}
S_{\textrm{IF}}[a^+,a^-]=-\frac{i}{2}\textrm{tr} \ln G.
\label{2.23}
\end{equation}

The matrix $G$ can be computed perturbatively. Following
Refs.~\cite{campos94,campos97,calzetta97c} we define $A=A^0+V$, where
the matrix $V$ includes the time-dependent interaction with
$V_{++}=-\nu (a^+)^2 R^+$, $V_{--}=\nu (a^-)^2 R^-$. Then up to second
order in $\nu$, $G=G^0(1-VG^0+VG^0VG^0+\dots)$ where $G^0$ is the
Minkowski $2\times2$ CTP propagator with $G^0_{++}=\Delta_F$,
$G^0_{--}=-\Delta_D$, $G^0_{+-}=-\Delta^+$ and $G^0_{-+}=\Delta^-$,
and where $\Delta_F$ and $\Delta_D$ are, respectively, the Feynman and
Dyson propagators and $\Delta^\pm$ are the Wightman functions:
\begin{eqnarray}
&&\Delta_{F/D}(x)=-\int\frac{d^n p}{(2\pi)^n}\frac{e^{ip\cdot x}}{p^2
\mp i\epsilon},\nonumber\\
&&\Delta^{\pm}(x)=\pm 2\pi i \int\frac{d^n p}{(2\pi)^n}e^{ip\cdot x}\delta(p^2)\theta(\mp p^0).
\label{2.24}
\end{eqnarray}

Substituting into Eq.~(\ref{2.23}) we have (up to second order in
$\nu$)
\begin{eqnarray}
S_{\textrm{IF}}[a^+,a^-]&=&-\frac{i}{2}\textrm{tr}\ln G^0+ \frac{i}{2}\textrm{tr} (V_{++}\Delta_F)- \frac{i}{2}\textrm{tr} (V_{--}\Delta_D)
\nonumber\\&&-\frac{i}{4}\textrm{tr} (V_{++}\Delta_F V_{++}\Delta_F)-\frac{i}{4}\textrm{tr} (V_{--}\Delta_DV_{--}\Delta_D)\nonumber\\
&&+ \frac{i}{2}\textrm{tr} (V_{++}\Delta^+ V_{--}\Delta^-).
\label{2.25}
\end{eqnarray}
The first three terms do not contribute to the dynamical equations for
$a(\eta)$, the first term is independent of $a$, and the second and
third terms are tadpoles which are identically zero in dimensional
regularization \cite{leibbrandt75}, so that there is no linear term in
$\nu$ in the effective action. The fourth and fifth terms involve the
product of Feynman and Dyson propagators and need regularization,
whereas the last term is finite. Following closely
Refs.~\cite{campos94, campos97} we get, after expanding in powers of
$(n-4)$, that the real part of $S_{\textrm{IF}}$ in $n$ dimensions is
\begin{eqnarray}
\textrm{Re}S_{\textrm{IF}}[a^+,a^-]=-\frac{9\nu^2{\cal V}}{8\pi^2} && \!\!\!\! \left\{
\frac{1}{n-4}\int d\eta \Delta\left(\left(\frac{\ddot a}{a}\right)^2\right)+ \frac{1 }{3}\int d\eta \Delta \left(\frac{\ddot a}{a}\left[3\left(\frac{\dot a}{a}\right)^2 +2\frac{\ddot a}{a}\right]\right)\right. \nonumber\\
&&\left.      +2\int\int d\eta d\eta^\prime \Delta\left(\frac{\ddot a}{a}(\eta)\right)\tilde{H}(\eta-\eta^\prime) \, \Sigma \! \left(\frac{\ddot a}{a}(\eta^\prime)\right) \right\}  + O(n-4),
\label{2.26}
\end{eqnarray}
where we have used the difference and semisum notations $\Delta
(f)\equiv f^+-f^-$ and $\Sigma (f) \equiv(f^+ +f^-)/2$, respectively,
and the kernel $\tilde{H}(\eta-\eta^\prime)$ is given by
\begin{equation}
\tilde{H}(\eta-\eta^\prime)= \int \frac{d\omega}{2\pi}e^{-i\omega (\eta-\eta^\prime)}\left(\ln |\omega| +\frac{i\pi}{2}\textrm{sign}(-\omega) - \frac{1}{2}(2+\ln 4\pi -\gamma) \right),
\label{2.27a}
\end{equation}
where $\gamma$ is the Euler-Mascheroni constant.  Note that although
the argument of the logarithm is not dimensionless, when combining the
influence action with the counterterms in the bare gravitational
action, the contribution involving the renormalization scale $\mu$
will finally render the argument of the logarithm dimensionless, as we
will see below.
There is also an imaginary part in $S_{\textrm{IF}}$, but it does not
contribute to the expectation value (\ref{2.17}), and thus to the
semiclassical equation for $a$, because it depends quadratically on
the difference variable $\Delta (\ddot a/a)$. This means that when
functionally deriving with respect to $a^+$ and then taking
$a^+=a^-=a$ to get the expectation value, the imaginary contribution
vanishes, as it should. The role of the imaginary part of
$S_{\textrm{IF}}$ is related to the so-called noise kernel, which
accounts for the fluctuations of the stress tensor and allows to go
beyond the semiclassical equations, which capture only the averaged
value of the stress tensor. The noise kernel plays a key role in
stochastic gravity, see Refs.~\cite{martin99a,martin99b,hu03a,hu04a}
for the general theory and
Refs.~\cite{calzetta94,hu95a,hu95b,campos96,calzetta97c} for
cosmological applications.

As explained in the appendix, 
the dynamical equations for the gravitational field can be derived
from the so-called CTP effective action, $\Gamma[a^+,a^-]$, the two
ingredients of which are the gravitational action $S_\mathrm{g}[a]$,
given by Eq.~(\ref{2.7}), and the influence action
$S_{\textrm{IF}}[a^+,a^-]$. Specializing Eq.~(\ref{2.11a}) to the
conformally flat metric in Eq.~(\ref{2.1}), the regularized CTP
effective action becomes
\begin{equation}
\Gamma[a^+,a^-] = S_\mathrm{g}[a^+]-S_\mathrm{g}[a^-]+ S_{\textrm{IF}}[a^+,a^-].
\label{2.28a}
\end{equation}
The real part of the regularized influence action is given by
Eq.~(\ref{2.26}) and diverges for $n=4$. Thus, we need to add
appropriate local covariant counterterms, which we denoted by
$S_\mathrm{g}^\mathrm{c}[a]$, to the bare gravitational action. For a
massless field only counterterms quadratic in the curvature, as
explicitly given by Eq.~(\ref{2.14b}), are needed. The integrand of
the first term on the right-hand side of Eq.~(\ref{2.14b}), which is
independent of $\nu^2$, is proportional to the square of the Weyl
tensor in $n=4$ dimensions. For a conformally flat metric like the
metric (\ref{2.1}) that we are considering here, the Weyl tensor
vanishes and thus this term vanishes in four dimensions. However, it
plays a crucial role in the trace anomaly \cite{birrell94}. In fact,
expanding in powers of $(n-4)$ we have
\begin{equation}
\mu^{n-4} \int d^nx\sqrt{-g}\left(R_{\mu\nu\rho\sigma}R^{\mu\nu\rho\sigma}-R_{\mu\nu}R^{\mu\nu}\right)=
-{\cal V}(n-4)\int d\eta \left[ 3\left(\frac{\ddot a}{a}\right)^2-
\left(\frac{\dot a}{a}\right)^4\right]+ O\left((n-4)^2\right),
\label{2.29a}
\end{equation}
which is of order $O(n-4)$ and therefore gives a finite contribution
when multiplied by the divergent $(n-4)^{-1}$ factor. On the other
hand, due to the $O(n-4)$ dependence there will be no contribution
proportional to the parameter $\alpha$ from Eq.~(\ref{2.14b}), as
expected since the tensor $A_{\mu\nu}$ in Eq.~(\ref{2.8}) vanishes for
a conformally flat metric. As for the second term on the right-hand
side of Eq.~(\ref{2.14b}), we have
\begin{equation}
\mu^{n-4} \int d^nx\sqrt{-g}R^2={\cal V}\int d\eta\left\{ 36
\left(\frac{\ddot a}{a}\right)^2
+
12(n-4) \left\{3\left(\frac{\ddot a}{a}\right)^2\ln (a \, \mu) + \frac{\ddot a}{a}\left[3\left(\frac{\dot a}{a}\right)^2 +2\frac{\ddot a}{a}\right]\right\}\right\}+ O\left((n-4)^2\right).
\label{2.29b}
\end{equation}
When multiplied by $(n-4)^{-1}$, the first term of this expansion
cancels out the $(n-4)^{-1}$ divergence of Eq.~(\ref{2.26}).

Finally, we can add the regularized counterterms of Eq.~(\ref{2.14b})
with the particular values in Eqs.~(\ref{2.29a}) and (\ref{2.29b}) to
the Einstein-Hilbert action (including the cosmological constant) to
obtain the total bare gravitational action $S_\mathrm{g}[a]$. Together
with the regularized influence action, whose real part is given by
Eq.~(\ref{2.26}), it gives the regularized CTP effective action
$\Gamma[a^+,a^-]$. We can then take the limit $n\to 4$ to obtain the
four-dimensional effective action in terms of the renormalized
gravitational action and influence action:\begin{equation}
  \Gamma[a^+,a^-] =
  S^{\mathrm{ren}}_\mathrm{g}[a^+]-S^{\mathrm{ren}}_\mathrm{g}[a^-] +
  S^{\mathrm{ren}}_{\textrm{IF}}[a^+,a^-].
\label{2.28b}
\end{equation}
The result for the renormalized real part of the influence action in
four dimensions is:
\begin{eqnarray}
\textrm{Re}S^{\textrm{ren}}_{\textrm{IF}}[a^+,a^-] = {\cal V}\int d\eta && \!\!\! \left\{
-\frac{1}{2880\pi^2}\Delta\left( 3\left(\frac{\ddot a}{a}\right)^2
-\left(\frac{\dot a}{a}\right)^4\right)\right.\nonumber\\
&&\left. +\frac{9\nu^2}{8\pi^2}\left[\Delta \left(\left(\frac{\ddot a}{a}\right)^2\ln a\right)- 2\int d\eta^\prime\Delta \left(\frac{\ddot a}{a}(\eta)\right)H(\eta-\eta^\prime;\bar{\mu}) \, \Sigma \! \left( \frac{\ddot a}{a}(\eta^\prime)\right)\right]\right\},
\label{2.30}
\end{eqnarray}
where we have incorporated the renormalization scale $\bar{\mu}= \mu
\exp[(2+\ln 4\pi -\gamma)/2]$ in the new kernel
\begin{equation}
H(\eta-\eta^\prime;\bar{\mu})= \int \frac{d\omega}{2\pi}e^{-i\omega (\eta-\eta^\prime)}\left(\ln \frac{|\omega|}{\bar{\mu}} +\frac{i\pi}{2}\textrm{sign}(-\omega) \right).
\label{2.27b}
\end{equation}
From the action in Eq.~(\ref{2.30}) one can obtain the renormalized
expectation value of the trace of the stress tensor (remember that the
imaginary part of the influence action plays no role in
that). Similarly, by taking the functional derivative of the CTP
effective action $\Gamma[a^+,a^-]$ with respect to $a^+$ and then
equating $a^+=a^-=a$ we obtain the trace of the semiclassical Einstein
equation.  Note that the CTP effective action, given by
Eq.~(\ref{2.28a}) or equivalently by Eq.~(\ref{2.28b}), is
renormalization-group invariant, \emph{i.e.}, it is independent of the
renormalization group scale $\mu$, and so are the physical predictions
that one can derive from it. The dependence on $\mu$ in
$\textrm{Re}S^{\textrm{ren}}_{\textrm{IF}}[a^+,a^-]$, which gives rise
to a local term of the form $(9\nu^2 / 8\pi^2) \ln \mu \, \Delta \!
\left( (\ddot a / a)^2 \right)$, is exactly compensated by the the
dependence on $\mu$ of the renormalized parameter $\beta$ multiplying
the $R^2$ term in the renormalized gravitational action. This can be
traced back to the fact that the bare parameter $\beta_\mathrm{B}$ is
independent of $\mu$, as explained in the appendix. 

We close this section by mentioning an alternative (but entirely
equivalent) method of calculating the influence action provided in
Ref.~\cite{roura99a}. The approach, which is based on decomposing the
field in spatial Fourier modes, computing the unitary evolution
operator for each mode perturbatively in the interaction picture, and
summing over all the modes at the end, can be useful when considering
more general initial states at a finite initial time $\eta_\mathrm{i}$
which are not necessarily of the form given by Eq.~(\ref{2.15}).

\section{The expectation value of the stress tensor}
\label{sec4}

\subsection{The trace}
\label{sec4.1}

Functionally differentiating with respect to $a^+$ the expression for
the influence action given by Eq.~(\ref{2.30}) and using
Eq.~(\ref{2.17}), we obtain the expectation value of the trace of the
stress tensor:
\begin{eqnarray}
\langle \hat{T}^\mu_\mu\rangle = \frac{1}{a^3} && \! \!  \! \! \! \left\{
-\frac{6}{2880\pi^2} \left[\frac{d^2}{d\eta^2}\left(\frac{\ddot a}{a^2}\right)
-\frac{\ddot a^2}{a^3}\right]
-\frac{4}{2880\pi^2}\left[\frac{d}{d\eta}\left(\frac{\dot a^3}{a^4}\right)+\frac{\dot a^4}{a^5}\right] \right. \nonumber\\
&&\left.
+\frac{9\nu^2}{4\pi^2}\left[\frac{d^2}{d\eta^2}\left(\frac{\ddot a}{a^2}\ln a\right)-\frac{\ddot a^2}{a^3}\left(\ln a-\frac{1}{2}\right)-\frac{d^2}{d\eta^2}\left(\frac{1}{a}\,\kappa\!\left[\frac{\ddot a}{a};\eta\right)\right)+\frac{\ddot a}{a^2}\,\kappa\!\left[\frac{\ddot a}{a};\eta\right)\right]\right\},
\label{4.1}
\end{eqnarray}
where
$\kappa[f;\eta)=\int^{\infty}_{-\infty}d\eta^\prime H(\eta-\eta^\prime
;\bar{\mu})f(\eta^\prime)$. One can easily check that the usual result
for the trace anomaly is obtained in the conformal limit $\nu\to
0$. Indeed, taking into account that in four dimensions
\begin{equation}
\Box R=-\frac{6}{a^3}\left[\frac{d^2}{d\eta^2}\left(\frac{\ddot a}{a^2}\right)-\frac{\ddot a^2}{a^3}\right]
\label{4.2}
\end{equation}
and
\begin{equation}
R_{\mu\nu\rho\sigma}R^{\mu\nu\rho\sigma}-R_{\mu\nu}R^{\mu\nu}=-\frac{4}{a^3}\left[\frac{d}{d\eta}\left(\frac{\dot a^3}{a^4}\right)+\frac{\dot a^4}{a^5}\right],
\label{4.3}
\end{equation}
Eq.~(\ref{4.1}) can be rewritten as
\begin{equation}
\langle \hat{T}^\mu_\mu\rangle =
\frac{1}{2880\pi^2}
\Box R+\frac{1}{2880\pi^2}\left(R_{\mu\nu\rho\sigma}R^{\mu\nu\rho\sigma}-R_{\mu\nu}R^{\mu\nu}\right)+O(\nu^2),
\label{4.4}
\end{equation}
which coincides with the trace anomaly \cite{birrell94} for a massless
conformal scalar field when $\nu=0$. Note that the counterterm in
Eq.~(\ref{2.29a}), when multiplied by the divergent factor
$(n-4)^{-1}$, plays a key role for this result.

Eq.~(\ref{4.1}) has a non-local term which includes the functional
$\kappa[\ddot a/a;\eta)$ and its first and second derivatives. Let us
examine this non-local part in some detail. The Fourier transform in
Eq.~(\ref{2.27b}) can be computed [see Eq.~(VII.7.18) in
Ref.~\cite{schwartz66}] to yield
\begin{equation}
H(\eta-\eta^\prime ;\bar{\mu})=-{\cal P}\! f \frac{\theta(\eta-\eta^\prime)}{\eta-\eta^\prime}-(\gamma+\ln \bar{\mu}) \, \delta(\eta-\eta^\prime),
\label{4.5}
\end{equation}
where ${\cal P}\! f$ stands for Hadamard's finite part prescription,
and $\gamma$ is the Euler-Mascheroni constant. This prescription means
that
\begin{equation}
\kappa[f;\eta)=-\lim_{\epsilon\to 0^+}\left\{\int^{\eta-\epsilon}_{-\infty} \frac{d\eta^\prime}{\eta-\eta^\prime} f(\eta^\prime)+(\ln\epsilon+\ln\bar{\mu}+\gamma)f(\eta)\right\}.
\label{4.6}
\end{equation}
Since the evolution from $\eta = -\infty$ to $\eta_\mathrm{i}$ can be
regarded as an auxiliary way to generate the initial state of the
quantum field, which is determined by the scale factor $a(\eta)$ for
times $\eta\le\eta_\mathrm{i}$ and denoted earlier by $a_\Psi (\eta)$,
it is convenient to define $v(\eta)\equiv (\ddot
a_\Psi/a_\Psi)(\eta)$. Therefore, for $\eta\le\eta_\mathrm{i}$ we have
$\kappa [\ddot a_\Psi/a_\Psi;\eta)=\kappa[v;\eta)$, whereas for
$\eta>\eta_\mathrm{i}$ the integral in Eq.~(\ref{4.6}) can be
separated into two parts:
\begin{equation}
\kappa\! \left[\frac{\ddot a}{a};\eta \right)=\kappa_1\! \left[\frac{\ddot a}{a};\eta \right) +\kappa_2 \left[v;\eta \right),
\label{4.8}
\end{equation}
where
\begin{equation}
\kappa_1\!\left[\frac{\ddot a}{a};\eta \right)=-\lim_{\epsilon\to 0^+}\left\{\int^{\eta-\epsilon}_{\eta_\mathrm{i}} \frac{d\eta^\prime}{\eta-\eta^\prime} \frac{\ddot a}{a}(\eta^\prime) +(\ln\epsilon+\ln\bar{\mu}+\gamma)\frac{\ddot a}{a}(\eta)\right\},
\label{4.9}
\end{equation}
which involves a time integration only after $\eta_\mathrm{i}$,
and
\begin{equation}
\kappa_2 [v;\eta)=-\int^{\eta_\mathrm{i}}_{-\infty} \frac{d\eta^\prime}{\eta-\eta^\prime}v(\eta^\prime),
\label{4.10}
\end{equation}
which involves a time integration only before $\eta_\mathrm{i}$.
Following this notation one can also separate the trace in
Eq.~(\ref{4.1}) for $\eta>\eta_\mathrm{i}$ into two parts: $\langle
\hat{T}^\mu_\mu\rangle=\langle \hat{T}^\mu_\mu\rangle_1+\langle
\hat{T}^\mu_\mu\rangle_2$, where the first term involves
$\kappa_1[\ddot a/a;\eta)$ and the second one involves $\kappa_2
[v;\eta)$.  From Eqs.~(\ref{2.5}) and (\ref{2.15}), and taking into
account that in four dimensions $a^2R=6\ddot a/a$, we see that
$v(\eta)$ completely determines the initial state of the quantum
field. Thus, $\langle \hat{T}^\mu_\mu\rangle_1$ is state-independent,
whereas $\langle \hat{T}^\mu_\mu\rangle_2$ contains all the dependence
on the initial state. In fact, the former will appear in the trace
even if the initial state does not have the form given by
Eq.~(\ref{2.15}).

\subsection{The initial conditions}
\label{sec4.2}

We can now see that the state-independent part of the trace of the
stress tensor, namely $\langle \hat{T}^\mu_\mu\rangle_1$, diverges in
the limit $\eta\to\eta_\mathrm{i}^+$, corresponding to the initial
time. Let us consider the definition of $\kappa_1[\ddot a/a;\eta)$ in
Eq.~(\ref{4.9}) and Taylor expand $\ddot a/a$ around $\eta'=\eta$ so
that $(\ddot a/a)(\eta^\prime)=(\ddot a/a)(\eta)+O(\eta-\eta^\prime)$;
we then have
\begin{equation}
\int^{\eta-\epsilon}_{\eta_\mathrm{i}} \frac{d\eta^\prime}{\eta-\eta^\prime} \frac{\ddot a}{a}(\eta^\prime)=-\frac{\ddot a}{a}(\eta)\ln\epsilon+\frac{\ddot a}{a}(\eta)\ln(\eta-\eta_\mathrm{i})+O(\eta-\eta_\mathrm{i}),
\label{4.11}
\end{equation}
which implies that
\begin{equation}
\kappa_1\!\left[\frac{\ddot a}{a};\eta\right)=-\frac{\ddot a}{a}(\eta)\left[\ln(\eta-\eta_\mathrm{i})+\ln\bar{\mu}+\gamma\right]+O(\eta-\eta_\mathrm{i}),
\label{4.12}
\end{equation}
so that $\kappa_1[\ddot a/a;\eta)$ is finite for all
$\eta>\eta_\mathrm{i}$, but diverges when
$\eta\to\eta_{i}^+$. Moreover, in addition to $\kappa_1[\ddot
a/a;\eta)$, $\langle \hat{T}^\mu_\mu\rangle_1$ also contains the first
and second time derivatives of $\kappa_1$. Obviously upon derivation
Eq.~(\ref{4.12}) becomes more singular at $\eta\to\eta_\mathrm{i}^+$,
and in general the divergencies of the three singular terms will not
cancel out. Therefore, $\langle \hat{T}^\mu_\mu\rangle_1\to \infty$ in
the limit $\eta\to\eta_\mathrm{i}^+$.

Thus, $\langle \hat{T}^\mu_\mu\rangle$ will diverge at the initial
time unless we choose an initial state that cancels the previous
divergencies. This partly motivates the class of initial states we
consider in this paper.  Taylor expanding $v$ around
$\eta'=\eta_\mathrm{i}$, and following a procedure analogous to the
one that led to Eq.~(\ref{4.11}) [notice that condition (\ref{2.6})
ensures that $\kappa_2[v;\eta)$ is well behaved in its lower limit],
one can check that
\begin{equation}
\kappa_2[v;\eta)=v(\eta_\mathrm{i})\ln(\eta-\eta_\mathrm{i})+\textrm{finite terms}
\label{4.13} ,
\end{equation}
where by ``finite terms'' we mean terms which are finite in the limit
$\eta \to \eta_\mathrm{i}$. Therefore, $\kappa[\ddot
a/a;\eta)=\kappa_1[\ddot a/a;\eta)+\kappa_2[v;\eta)$ is finite at
$\eta_\mathrm{i}$ provided that
$v(\eta_\mathrm{i})=\lim_{\eta\to\eta_\mathrm{i}^+}(\ddot a/a)$. If
this condition is satisfied, it is easy to see from Eqs.~(\ref{4.9})
and (\ref{4.10}) that $\frac{d}{d\eta}\kappa[\ddot
a/a;\eta)=\kappa[\frac{d}{d\eta}(\ddot a/a);\eta)$. Thus, by iterating
the same argument we find that the conditions that an initial state of
the class (\ref{2.15}) must satisfy in order to avoid initial time
divergencies of $\langle \hat{T}^\mu_\mu\rangle$ are
\begin{equation}
v(\eta_\mathrm{i})=\lim_{\eta\to\eta_\mathrm{i}^+}\left(\frac{\ddot a}{a}\right),\qquad\dot v(\eta_\mathrm{i})=\lim_{\eta\to\eta_\mathrm{i}^+}\frac{d}{d\eta}\left(\frac{\ddot a}{a}\right),\qquad\ddot v(\eta_\mathrm{i})=\lim_{\eta\to\eta_\mathrm{i}^+}\frac{d^2}{d\eta^2}\left(\frac{\ddot a}{a}\right).
\label{4.14}
\end{equation}
In other words, the cosmological scale factors before and after the
initial time must be matched with continuity up to the fourth
derivative. Eqs.~(\ref{4.14}) have been presented as conditions on the
preparation of the initial state given a scale factor $a(\eta)$ for
$\eta \ge \eta_\mathrm{i}$. Alternatively, one can regard these
equations as initial conditions on the scale factor given some initial
state defined by Eq.~(\ref{2.15}). Henceforth we will assume that
these conditions are satisfied.

\subsection{The energy density}
\label{sec4.3}

Once the expectation value of the trace of the stress tensor is known,
the $00$ component of the stress tensor may be obtained from
Eq.~(\ref{2.19}). The local part of $\dot a a^3 \langle
\hat{T}^\mu_\mu\rangle$ is a derivative, so it gives rise to local
terms in $a^2\langle \hat{T}_{00}\rangle$. The result, for
$\eta>\eta_\mathrm{i}$, is
\begin{eqnarray}
a^2\langle \hat{T}_{00}\rangle&=&\frac{6}{2880\pi^2}\left[\dot a \frac{d}{d\eta}\left(\frac{\ddot a}{a^2}\right)-\frac{1}{2}\left(\frac{\ddot a}{a}\right)^2\right]+\frac{3}{2880\pi^2}\left(\frac{\dot a}{a}\right)^4\nonumber\\&&-\frac{9\nu^2}{4\pi^2}\left\{\ln a\left[\dot a \frac{d}{d\eta}\left(\frac{\ddot a}{a^2}\right)-\frac{1}{2}\left(\frac{\ddot a}{a}\right)^2\right]+\frac{\dot{a}^2\ddot a}{a^3}+T[a,v]\right\},
\label{4.15}
\end{eqnarray}
where $T$, which includes the non-local part of this expectation value
and comes from integrating the last two terms in Eq.~(\ref{4.1}), is
defined by
\begin{equation}
T[a,v;\eta) = \int^{\eta}_{-\infty} d\eta' \, \dot{a} \left[
- \frac{d^2}{d\eta^{\prime\,2}}\left(\frac{1}{a}\,\kappa\!\left[\frac{\ddot a}{a};\eta'\right)\right)
+ \frac{\ddot a}{a^2}\,\kappa\!\left[\frac{\ddot a}{a};\eta'\right) \right]
\label{4.16} .
\end{equation}
Making use of the ``smoothness'' of the scale factor at
$\eta_\mathrm{i}$ assumed above, and following the previous notation
for the separation of the non-local terms, we can again write
\begin{equation}
T[a,v]=T_1[a]+T_2[a,v],
\label{4.17}
\end{equation}
where we split $T[a,v]$ into two parts: one independent of $v$ and a
second one which contains the entire dependence on $v$. This is
achieved by splitting the integral in Eq.~(\ref{4.16}) as well as the
functional $\kappa$. The first part is given by
\begin{equation}
T_1[a;\eta)=\int^{\eta}_{\eta_\mathrm{i}}d\eta^\prime\left(\frac{\dot a}{a}\right)\!(\eta^\prime)\left\{2\frac{d}{d\eta^\prime}\left(\frac{\dot a}{a}\right)\kappa_1\!\left[\frac{\ddot a}{a};\eta^\prime\right)+2\left(\frac{\dot a}{a}\right)\!(\eta^\prime)\;\kappa_1\!\left[\frac{d}{d\eta''}\left(\frac{\ddot a}{a}\right);\eta^\prime\right)-\kappa_1\!\left[\frac{d^2}{d\eta^{\prime\prime\,2}}\left(\frac{\ddot a}{a}\right);\eta^\prime\right)\right\}.
\label{4.18}
\end{equation}
This is obtained from Eq.~(\ref{4.16}) by explicitly applying the
second-order derivative $d^2/d\eta^{\prime\,2}$ and taking into
account that when the conditions in Eqs.~(\ref{4.14}) hold, the
derivatives acting on the functional $\kappa$ can be taken inside and
applied to the argument. All this is done before splitting the
integral and the functional $\kappa$.  Similarly, the second part in
Eq.~(\ref{4.17}) can be written as
\begin{eqnarray}
T_2[a,v;\eta)&=&\int^{\eta}_{\eta_\mathrm{i}}d\eta^\prime\left(\frac{\dot a}{a}\right)\!(\eta^\prime)\left\{2\frac{d}{d\eta^\prime}\left(\frac{\dot a}{a}\right)\kappa_2[v;\eta^\prime)+2\left(\frac{\dot a}{a}\right)\!(\eta^\prime)\;\kappa_2[\dot v;\eta^\prime)-\kappa_2[\ddot v;\eta^\prime)\right\}\nonumber\\&&+\left(\frac{\dot a}{a}\right)^{\!2}\!(\eta_\mathrm{i})\;\kappa[v;\eta_\mathrm{i})-\left(\frac{\dot a}{a}\right)\!(\eta_\mathrm{i})\;\kappa[\dot v;\eta_\mathrm{i})+\int^{\eta_\mathrm{i}}_{-\infty}d\eta^\prime v(\eta^\prime)\kappa[\dot v;\eta^\prime).
\label{4.19}
\end{eqnarray}
The different form of the last three terms, which correspond to the
integral from $-\infty$ to $\eta_\mathrm{i}$, is because we made use
of the following equivalent expression for the last two terms in
Eq.~(\ref{4.1}) times $(-\dot{a} a^3)$:
\begin{equation}
-\dot a\frac{d^2}{d\eta^{\prime\,2}}\left(\frac{1}{a}\kappa\!\left[\frac{\ddot a}{a};\eta'\right)\right)
+ \dot{a} \frac{\ddot a}{a^2}\,\kappa\!\left[\frac{\ddot a}{a};\eta'\right)
= - \frac{d}{d\eta'} \left( \dot a\frac{d}{d\eta'}\left(\frac{1}{a}
\kappa\!\left[\frac{\ddot a}{a};\eta'\right)\right)\right)
+ \frac{\ddot a}{a} \frac{d}{d\eta'} \kappa\!\left[\frac{\ddot a}{a};\eta' \right)
\label{4.20} .
\end{equation}
In addition to being convenient later on, we did this so that it
became manifest that $T_2[a,v]$ depends on $a_\Psi$ only through $v$.
From Eqs.~(\ref{4.18})-(\ref{4.19}) it is clear that all the
dependence of the energy density on the initial state of the quantum
field is included in $T_2[a,v]$. We should, however, remember that
conditions (\ref{4.14}) have been used so that the divergencies from
the evolution after $\eta_\mathrm{i}$ and those from the initial state
would cancel out. Consequently, the integrand on the right-hand side
of Eq.~(\ref{4.18}) and the integrand of the first integral in
Eq.~(\ref{4.19}) only exhibit logarithmic divergences when $\eta' \to
\eta_\mathrm{i}$ which would cancel out when adding $T_1$ and
$T_2$. In fact, they give a finite contribution even separately since,
being logarithmic, they are finite upon integration.

The energy density that we have just computed is valid for any scale
factor at $\eta\geq\eta_\mathrm{i}$ and an initial state corresponding
to any regular function $v(\eta)$ which satisfies conditions
(\ref{2.6}) and (\ref{4.14}). It can be included in the semiclassical
Friedmann equation in the presence of any other classical source. In
this paper we concentrate on the case where the classical source is a
cosmological constant.

\section{The semiclassical Friedmann equation}

\label{sec5}

\subsection{The equation}
\label{sec5.1}

Next, we proceed to analyze the evolution of the scale factor after
the initial time $\eta_\mathrm{i}$, when it is driven by a
cosmological constant and a free massless scalar field non-conformally
coupled to the curvature. We will use the $00$ component of the
semiclassical Einstein equation (\ref{2.8}), namely, $G_{00}=-\Lambda
g_{00} - \beta B_{00} +\kappa\langle
\hat{T}_{00}\rangle_\mathrm{ren}$. Note that we have already taken
into account that $A_{\mu\nu}=0$ in our conformally flat spacetime. On
the other hand, evaluating $B_{\mu\nu}$, as given by Eq.~(\ref{2.9b}),
for the RW metric (\ref{2.1}), we get
\begin{equation}
B_{00}= - \frac{72}{a^2} \left[\dot a \frac{d}{d\eta}\left(\frac{\ddot a}{a^2}\right)
-\frac{1}{2}\left(\frac{\ddot a}{a}\right)^2\right]
\label{5.0},
\end{equation}
which can also be obtained by using Eq.~(\ref{2.29b}) for $n=4$ and
noticing that the term proportional to $\Delta (\ddot{a}/a)$ gives
rise to the first term on the right-hand side of
Eq.~(\ref{4.15}). Taking into account that $G_{00}=3(\dot a/a)^2$ for
the metric (\ref{2.1}) and using Eq.~(\ref{4.15}) we obtain the
following expression for the semiclassical Friedmann equation:
\begin{eqnarray}
\dot a^2 = H^2a^4 + \frac{l_\mathrm{p}^2}{3}&&\!\!\!\! \left\{-\tilde{\beta}\left[\dot a \frac{d}{d\eta}\left(\frac{\ddot a}{a^2}\right)-\frac{1}{2}\left(\frac{\ddot a}{a}\right)^2\right]+\frac{3}{360\pi}\left(\frac{\dot a}{a}\right)^4 \right. \nonumber\\
&& \left. -\frac{18\nu^2}{\pi}\left\{\ln a\left[\dot a \frac{d}{d\eta}\left(\frac{\ddot a}{a^2}\right)-\frac{1}{2}\left(\frac{\ddot a}{a}\right)^2\right]+\frac{\dot{a}^2\ddot a}{a^3}+T[a,v]\right\}\right\},
\label{5.1}
\end{eqnarray}
where $\tilde{\beta} = 576 \pi \beta - 1/(60 \pi)$, and we have
introduced the Hubble constant $H\equiv(\Lambda/3)^{1/2}$ and the
Planck length $l_\mathrm{p}=\sqrt{G}=(\kappa/8\pi)^{1/2}$. These two
constants introduce two different time-scales into the problem, the
Hubble time $H^{-1}$ and the Planck time $t_\mathrm{p}=l_\mathrm{p}$,
which we will assume to be well separated, namely, $H^{-1}\gg
l_\mathrm{p}$.  It should also be emphasized that the semiclassical
Friedmann equation (\ref{5.1}) is invariant under conformal-time
translations in the following sense. We already know that $v(\eta)$,
with domain $(-\infty,\eta_\mathrm{i}]$, characterizes the initial
state of the quantum field, $|\Psi (\eta_\mathrm{i})\rangle$, when it
is of the form given by Eq.~(\ref{2.15}). The time translation of
$v(\eta)$, given by $v_\Delta(\eta)=v(\eta+\Delta)$, is defined in the
domain $(-\infty,\eta_\mathrm{i}-\Delta]$ and corresponds to an
initial state $|\Psi_\Delta(\eta_\mathrm{i}-\Delta)\rangle=|\Psi
(\eta_\mathrm{i})\rangle$. It is then easy to see from Eq.~(\ref{5.1})
that if $a(\eta)$ is a solution for some initial state $|\Psi
(\eta_\mathrm{i})\rangle$ characterized by $v(\eta)$, then
$a(\eta+\Delta)$ is a solution for the initial state
$|\Psi_\Delta(\eta_\mathrm{i}-\Delta)\rangle$, characterized by
$v_\Delta(\eta)$.

Our semiclassical Friedmann equation (\ref{5.1}), which is a
non-linear third-order integro-differential equation, looks like a
typical back-reaction equation. Because of the higher-order time
derivatives, those equations exhibit extra degrees of freedom which
usually translate into unphysical runaway-type solutions
\cite{horowitz80}. [Indeed, if we had proceeded analogously to what
was done in Ref.~\cite{cooper89} for QED, the Friedmann equation and
the equation for the trace would fix the third and fourth derivatives
of the scale factor at the initial time, given some freely specified
initial values for $a$, $\dot{a}$ and $\ddot{a}$. This is in contrast
to the classical case, where only $a$ (or alternatively $\dot{a}$) can
be specified independently at the initial time.]  To get rid of the
unphysical solutions several methods have been proposed. In some
methods one looks only for analytic solutions, in a suitable
perturbative parameter, in order to select physical solutions only
\cite{simon90,simon91,simon92}. In some other methods the equation
itself is changed in order to get rid of the higher-order derivatives;
this is the case of the so-called order reduction method
\cite{parker93}. See Ref.~\cite{flanagan96} for a review of the
advantages and the shortcomings of the different methods in the
context of semiclassical gravity.

Given how we have proceeded in this paper, it is more natural in our
case to interpret the point discussed in the previous paragraph as
follows. The solutions of our semiclassical Friedmann equation
(\ref{5.1}) must satisfy the the three initial conditions in
Eqs.~(\ref{4.14}). This means that given some initial state,
completely characterized by the function $v(\eta)$, the semiclassical
Friedmann equation together with the equation for the trace fix
$a(\eta_\mathrm{i})$ and $\dot{a}(\eta_\mathrm{i})$ since the second,
third and fourth derivatives at the initial time are fixed by
Eqs.~(\ref{4.14}). Thus, there is in general a unique solution
compatible with a given initial state.  Certain initial states,
however, give rise to solutions with characteristic time-scales of the
order of the Planck time (corresponding to exponential growth or
oscillatory behavior), which lie beyond the regime of validity of the
low-energy effective field theory approach that we have implicitly
been using. In those circumstances the higher-order corrections
involving terms with positive powers of the curvature and suppressed
by the corresponding powers of the Planck mass can no longer be
neglected and the low-energy expansion breaks down (since there are in
principle an infinite number of such terms).  We will consider only
situations with no Planckian features where higher-order corrections
are negligible. Hence, in the spirit of the effective field theory
approach (valid for $l_\mathrm{p} H \ll 1$), we will look for
perturbative solutions%
\footnote{A perturbative expansion may sometimes miss the right
  long-time behavior of the semiclassical solution. This can happen
  when the effect of the quantum corrections, although locally small,
  builds up over long times giving rise to substantial deviations from
  the classical solution. An example of such a situation is the
  evolution of a black hole spacetime when the back reaction of the
  emitted Hawking radiation is taken into account. One possibility in
  those cases is to modify the back-reaction equation using an
  order-reduction procedure and then solve the resulting equation
  non-perturbatively~\cite{hu07b}. In the cosmological case considered
  here one can argue that such an accumulation effect will not be
  present. That is because the classical cosmological constant implies
  a monotonous growing behavior for the unperturbed solution and the
  locally small effect due to the vacuum polarization of the quantum
  fields (much smaller than the classical cosmological constant)
  generates a perturbation which is always small compared to the
  unperturbed solution. Moreover, one can explicitly check \emph{a
    posteriori} that the deviation from the classical solution does
  not become significant at late times.}  in powers of $(l_\mathrm{p}
H)^2$ as
\begin{equation}
a(\eta) = a_0(\eta) + (l_\mathrm{p} H)^2 a_1(\eta)+O\bigl( (l_\mathrm{p} H)^4 \bigr),
\label{5.2}
\end{equation}
where $a_0(\eta)$ satisfies the classical Friedmann equation. From now
on we will use the subscript $0$ to indicate the classical unperturbed
values. Without loss of generality we can focus on the solutions
$a(\eta)$ with $a_0(\eta)=-1/H\eta$ defined for
$\eta_\mathrm{i}<\eta<0$ since any other well-behaved solution not
involving Planckian scales is connected to one of these by a time
translation. The function characterizing the initial state is given by
$v(\eta)=v_0(\eta)+O\bigl( (l_\mathrm{p} H)^2 \bigr)$, and the
conditions in Eqs.~(\ref{4.14}) imply
\begin{equation}
v_0(\eta_\mathrm{i})=\frac{2}{\eta_{i}^2}\, ,\qquad \dot v_0(\eta_\mathrm{i})=-\frac{4}{\eta_{i}^3}\, ,\qquad \ddot v_0(\eta_\mathrm{i})=\frac{12}{\eta_{i}^4}\, .
\label{5.3}
\end{equation}
Substituting the expression (\ref{5.2}) into the semiclassical
Friedmann equation (\ref{5.1}), we obtain an equation for the first
order perturbation $a_1(\eta)$:
\begin{equation}
\dot a_1=-\frac{2}{\eta}a_1+\frac{1}{H}\left\{\frac{1}{720\pi}\frac{1}{\eta^2}-\frac{3\nu^2}{\pi}\left[\frac{2}{\eta^2}+\eta^2 T[a_0,v_0]\right]\right\},
\label{5.4}
\end{equation}
which is a first-order differential equation.  Notice that there is no
dependence on the free parameter $\tilde{\beta}$. This is not
surprising if one takes into account that $\tilde{\beta}$ is the
coefficient of a term proportional to $\Box R$ in the trace [see
Eqs.~(\ref{4.1}) and (\ref{4.2})] and that for the classical
background $R_0=4\Lambda$ is a constant (where $R_0$ is the Ricci
scalar for the classical background).

Let us now compute the non-local term $T[a_0,v_0]$ in the previous
equation. After substitution of $a_0$ into Eq.~(\ref{4.18}), the
state-independent contribution of this term reads
\begin{equation}
T_1[a_0;\eta)=-\frac{3}{2\eta^4}+\frac{1}{\eta_{i}^2\eta^2}
+\frac{4}{\eta_{i}^3\eta}-\frac{7}{2\eta_{i}^4},
\label{5.5}
\end{equation}
which 
is independent of the arbitrary renormalization scale $\mu$. The
latter is again because when substituting the local part of
$\kappa_1[f;\eta')$ into Eq.~(\ref{4.18}), one obtains an integrand
proportional to $\Box R_0 = 0$; this point could also have been
anticipated from the fact that the term proportional to $\ln \mu$
should have the same form as the term proportional to $\tilde{\beta}$,
which has been found to vanish above.%
\footnote{Note that the results obtained in
  Refs.~\cite{espriu05,cabrer07} in a similar context exhibit a
  non-vanishing dependence on $\mu$. This should not be the case for
  the reasons given here.}

Substituting $a_0$ and $v_0$ into Eq.~(\ref{4.19}), we get the
following result for the state-dependent part of $T[a_0,v_0;\eta)$:
\begin{eqnarray}
T_2[a_0,v_0;\eta)&=&\int_{\eta_\mathrm{i}}^{\eta}d\eta^\prime\int_{-\infty}^{\eta_\mathrm{i}}\frac{d\eta^{\prime\prime}}{\eta^\prime-\eta^{\prime\prime}}\left\{\frac{2}{{\eta^\prime}^3}v_0(\eta^{\prime\prime})-\frac{2}{{\eta^\prime}^2}\dot v_0(\eta^{\prime\prime})-\frac{1}{\eta^\prime}\ddot v_0(\eta^{\prime\prime})\right\}\nonumber\\&& +\frac{1}{\eta_{i}^2}\kappa[v_0;\eta_\mathrm{i})+\frac{1}{\eta_\mathrm{i}}\kappa[\dot v_0;\eta_\mathrm{i})+\int_{-\infty}^{\eta_\mathrm{i}}d\eta^\prime v_0(\eta^\prime)\kappa[\dot v_0;\eta^\prime).
\label{5.6}
\end{eqnarray}
The conditions (\ref{5.3})
guarantee that the dependence on $\mu$ of the last three terms cancels
out. The integral over $\eta^\prime$ in the first term can be easily
computed.  Finally, adding up the two contributions $T_1$ and $T_2$ we
get
\begin{equation}
T[a_0,v_0;\eta)=-\frac{3}{2\eta^4}+\frac{A_\Psi}{\eta^2}+B_\Psi+\int_{-\infty}^{\eta_\mathrm{i}}d\eta^\prime\ln\left[\frac{\eta_\mathrm{i}(\eta-\eta^\prime)}{\eta(\eta_\mathrm{i}-\eta^\prime)}\right]\left[\frac{2}{{\eta^\prime}^3}v_0(\eta^\prime)-\frac{2}{{\eta^\prime}^2}\dot v_0(\eta^\prime)-\frac{1}{\eta^\prime}\ddot v_0(\eta^\prime)\right],
\label{5.7}
\end{equation}
where
\begin{equation}
A_\Psi=\frac{1}{\eta_{i}^2}+\int_{-\infty}^{\eta_\mathrm{i}}d\eta^\prime\frac{v_0(\eta^\prime)}{\eta^\prime}
\label{5.8}
\end{equation}
and
\begin{equation}
B_\Psi=\frac{1}{2\eta_{i}^4}-\frac{1}{\eta_{i}^2}\int_{-\infty}^{\eta_\mathrm{i}}d\eta^\prime\frac{v_0(\eta^\prime)}{\eta^\prime}+\frac{1}{\eta_{i}^2}\kappa[v_0;\eta_\mathrm{i})+\frac{1}{\eta_\mathrm{i}}\kappa[\dot v_0;\eta_\mathrm{i})+\int_{-\infty}^{\eta_\mathrm{i}}d\eta^\prime v_0(\eta^\prime)\kappa[\dot v_0;\eta^\prime).
\label{5.9}
\end{equation}

We have already pointed out that the first order equation (\ref{5.4})
includes neither the free parameter $\tilde{\beta}$ nor the arbitrary
renormalization mass scale $\mu$. Moreover, it does not depend on
$\alpha$ either since $A_{\mu\nu}=0$ for a conformally flat
spacetime. This means that the semiclassical Friedmann equation
(\ref{5.1}) is fully predictive to this first perturbative order for
the particular case of a de Sitter background that we are considering.

\subsection{The solution}
\label{sec5.2}

The equation (\ref{5.4}) for the function $a_1(\eta)$ is a first
order, linear differential equation, which can be easily solved in an
explicit form. The general solution of a differential equation of the
type $y'(x)=-(2/x)y+f(x)$ is $y(x)=k/x^2+(1/x^2)\int dx x^2f(x)$,
where $k$ is an arbitrary integration constant. Therefore, the
solution of Eq.~(\ref{5.4}) is:
\begin{equation}
a_1(\eta)=\frac{k}{H^2 \eta^2}+\left(\frac{1}{720\pi}-\frac{3\nu^2}{2\pi}\right)\frac{1}{H \eta}-\frac{3\nu^2}{\pi H}\left[\frac{A_\Psi}{3}\eta+\frac{B_\Psi}{5}\eta^3+\Gamma_\Psi(\eta)\right],
\label{5.10}
\end{equation}
where
\begin{equation}
\Gamma_\Psi(\eta)=\int_{-\infty}^{\eta_\mathrm{i}}d\eta^\prime \gamma(\eta,\eta^\prime)\left[\frac{2}{{\eta^\prime}^3}v_0(\eta^\prime)-\frac{2}{{\eta^\prime}^2}\dot v_0(\eta^\prime)-\frac{1}{\eta^\prime}\ddot v_0(\eta^\prime)\right],
\label{5.11}
\end{equation}
with
\begin{equation}
\gamma(\eta,\eta^\prime)=\frac{1}{5\eta^2}\left\{\eta^5\ln\left[\frac{\eta_\mathrm{i}(\eta-\eta^\prime)}{\eta(\eta_\mathrm{i}-\eta^\prime)}\right]-{\eta^\prime}^5\left[\frac{1}{4}\left(\frac{\eta}{\eta^\prime}\right)^4+\frac{1}{3}\left(\frac{\eta}{\eta^\prime}\right)^3+\frac{1}{2}\left(\frac{\eta}{\eta^\prime}\right)^2+\frac{\eta}{\eta^\prime}+\ln\left(1-\frac{\eta}{\eta^\prime}\right)\right]\right\}.
\label{5.12}
\end{equation}
The condition (\ref{2.6}), which $v_0(\eta)$ must also satisfy,
ensures that the integral in Eq.~(\ref{5.11}) is finite in its lower
limit. As for the upper limit, all the possible divergencies in the
integrand cancel out (moreover, even if they did not cancel out, they
would give a finite contribution when integrated since they are
logarithmic). Thus, $\Gamma_\Psi(\eta)$ is finite and, furthermore, it
vanishes at future infinity, \emph{i.e.}, when $\eta\to0^-$.

The solutions associated with the different possible values of the
arbitrary integration constant $k$ are related to each other by time
translations: $a\bigl(\eta+(l_\mathrm{p}H)^2\Delta
\bigr)=-1/H\eta+(l_\mathrm{p}H)^2[a_1(\eta)+\Delta/H\eta^2]+O\bigl(
(l_\mathrm{p} H)^4 \bigr)$. For the particular case $k=0$, and
provided that $H^{-1} \gg l_\mathrm{p}$, the first-order term in the
expansion (\ref{5.2}) is much smaller than the zeroth-order term for
all times $\eta_\mathrm{i}<\eta<0$, and one expects that the
higher-order corrections can be neglected [since they are suppressed
by an additional power of $(l_\mathrm{p}H)^2$] and the solution $a(\eta)$ can
be approximated by the first-order perturbative expansion. This
approximation is no longer suitable for solutions (initial states)
associated with other values of $k$, but we already know that they are
just time translations of the solutions with $k=0$.
Therefore, we can conclude that for reasonable states, namely, those
which do not involve Planckian scales (and up to time translations)
the scale factor after the initial time $\eta_\mathrm{i}$ is given by
\begin{equation}
a(\eta) = -\frac{1}{\tilde H\eta}-(l_\mathrm{p}H)^2\frac{3\nu^2}{\pi H}
\left[\frac{A_\Psi}{3}\eta+\frac{B_\Psi}{5}\eta^3+\Gamma_\Psi(\eta)\right]
+ O\bigl( (l_\mathrm{p} H)^4 \bigr),
\label{5.13}
\end{equation}
where the modified Hubble constant is
\begin{equation}
\tilde H=H\left[1+(l_\mathrm{p}H)^2 \left(\frac{1}{720\pi}
-\frac{3\nu^2}{2\pi}\right)\right].
\label{5.14}
\end{equation}
The result in Eq.~(\ref{5.13}) is valid for all times (after the
initial time). It is the sum of a corrected de Sitter solution plus
terms that vanish at future infinity $\eta\to 0^-$. The corrected de
Sitter term dominates at late times, regardless of the initial
state. Therefore, we conclude that de Sitter spacetime is stable under
spatially-isotropic perturbations in semiclassical gravity.

The semiclassical shift of the Hubble constant does not have a
definite sign, it depends on the coupling between the quantum field
and the curvature. For instance, while the shift is positive for the
conformal coupling case ($\nu=0$), when $\nu=-1/6$, which is expected
to mimic the effect of gravitons, this shift is negative, implying a
small time-independent screening of the cosmological constant. The
deviation from the conformal coupling, $\nu$, is what makes the
Hamiltonian (\ref{2.5}) time dependent through the term involving the
scale factor. This is what allows us to prepare a wide range of
initial states by evolution of the \emph{in}-vacuum with different
scale factors $a_\Psi(\eta)$. In fact, the effect of this time
dependence of the Hamiltonian can be interpreted as particle
creation. Since the term proportional to $\nu^2$ in Eq.~(\ref{5.14})
is negative, one can say that at late times the effect of the created
particles is to slow down the de Sitter expansion by a small amount,
whereas the other vacuum polarization term, already present in the
conformal case, has the opposite effect.

A particularly interesting initial state is the Bunch-Davies vacuum
$|\Psi_\mathrm{BD}(\eta_\mathrm{i})\rangle$, which is obtained by
evolution of the \emph{in}-vacuum from $-\infty$ to $\eta_\mathrm{i}$
according to the scale factor of de Sitter spacetime or, in other
words, with $v_0(\eta)=2/\eta^2$. In this case, one can check that
$A_\Psi=B_\Psi=\Gamma_\Psi(\eta)=0$, so the solution (\ref{5.13}) is
just the de Sitter solution with a semiclassically modified Hubble
parameter. Therefore,
the Bunch-Davies vacuum together with de Sitter spacetime with the
semiclassically modified Hubble constant constitute a self-consistent
solution of the semiclassical Friedmann equation, to which other
perturbed solutions tend at late times.

\section{Discussion}
\label{sec6}

In this paper we have computed the one-loop vacuum polarization for
massless non-conformal fields in a general spatially-flat RW
background. This has then been applied to studying the evolution of
spatially-isotropic perturbations around de Sitter spacetime when the
back reaction due to a massless non-conformal field is
self-consistently included, which corresponds to solving the equations
of semiclassical gravity for this case. There is a self-consistent
solution, associated to the Bunch-Davies vacuum for the quantum
fields, with an effective cosmological constant slightly shifted from
its classical value due to the vacuum polarization
effects. Furthermore, we have found that this solution is stable under
spatially-isotropic perturbations since the perturbed solutions tend
to it at late times. It should be stressed that our results are
independent of the particular value of the renormalization parameters
$\alpha$ and $\beta$, and therefore fully predictive (at the first
order in $l_\mathrm{p}^2$ at which we are working). It should also be
pointed out that our results seem to be at variance with those of
Refs.~\cite{espriu05,cabrer07}. 
In this respect, it is important to keep in mind that
one should consider the stability of the self-consistent solution with
the shifted effective cosmological constant rather than that of the
classical solution obtained in the absence of vacuum polarization
effects.
Furthermore, when comparing the perturbed solution to the self-consistent
de Sitter ``attractor'' it is important to take properly into account any relative
conformal time translation since that can give rise to a spurious growth in
time of their ratio. This can be clearly illustrated by comparing two
copies of the same self-consistent de Sitter solution with a relative
conformal time translation.

We were able to obtain explicit analytic results by using two
approximations. First, we considered a perturbative expansion in the
parameter $\nu$, which characterizes the deviation of the curvature
coupling parameter from the conformal case, and truncated the
expansion to quadratic order. Hence, our effective action is exact
through order $\nu^2$. Second, we introduced a perturbative expansion
in powers of $(l_\mathrm{p} H)^2$. Its purpose was to obtain a fairly
accurate description for phenomena involving length-scales much larger
than the Planck length while discarding spurious solutions involving
Planckian scales, where the effective field theory approach that we
have been using breaks down.  Truncating such a perturbative expansion
for the solutions (rather than doing so at the level of the equation
of motion and then solving it exactly) can sometimes miss the right
long-time behavior. However, as we discussed in Sec.~\ref{sec5}, this
should not be the case for the situation considered here.

There are a number of natural extensions or generalizations of our
work in this paper. First, one could consider the other possibility
for having weakly non-conformal fields, namely, the case of fields
with conformal coupling but with a small non-vanishing mass such that
$m^2 \ll H^2$. In that case one should be able to proceed analogously
to what we did here by computing the effective action perturbatively
in $m^2$ through order $m^4$ and then solving the equation of motion
through order $(l_\mathrm{p} H)^2$. Second, the case of strongly
non-conformal fields with a large mass $M^2 \gg H^2$ could be
explicitly calculated using an adiabatic (or WKB)
expansion~\cite{frolov84}. The effective action in that case could be
written as a local expansion of positive powers of curvature
invariants suppressed by the corresponding power of $M^2$, a form
which can be anticipated from local effective field theory arguments
based on power counting and taking into account the relevant
symmetries (diffeomorphism invariance in this case). On the other
hand, the case of strong non-conformal coupling to the curvature could
be treated by introducing a field-dependent conformal transformation
relating the original Jordan frame to the Einstein frame,%
\footnote{Any physical predictions derived in the Jordan and Einstein
  frames should be equivalent, at least at the classical level. In the
  quantum mechanical case one still expects such an equivalence for
  small perturbations of the metric and the scalar field around their
  mean values \cite{flanagan04}. In that case, however, it may be
  necessary to treat the perturbations of the metric and the scalar
  field on an equal footing, and the metric perturbations may need to
  be quantized.}  where the curvature scalar does not couple to the
scalar field \cite{futamase89}.

Third, even though the construction based on Eq.~(\ref{2.15}) can
generate a fairly wide family of squeezed Gaussian states by
considering a sufficiently general form of the auxiliary scale factor
$a_\Psi(\eta)$, other approaches are needed in order to deal with all
possible Gaussian states or even non-Gaussian ones. One possibility is
to make use of the method developed in Ref.~\cite{habib00} for a fixed
background spacetime, which is based on the construction of
fourth-order adiabatic vacuum states (one can show that the class of
states that we have considered here are compatible with their approach
and encompassed by the general class of states it can deal
with). However, even if we restrict ourselves to the subclass of
initial states generated by Eq.~(\ref{2.15}), the procedure can be
straightforwardly generalized to states with non-vanishing expectation
values of the field or its canonically conjugate momentum. Since we
are considering Gaussian states, this can be done by decomposing the
field as a sum of a classical part which characterizes the evolution
of the expectation value and satisfies the classical equation of
motion plus a field with vanishing expectation value whose initial
state can be generated by Eq.~(\ref{2.15}). This implies that when
solving the back-reaction equation in powers of $(l_\mathrm{p} H)^2$
one first needs to find the self-consistent solution for the
background scale factor and the classical configuration of the field
(corresponding to the evolution of its expectation value), and then
solve for the perturbation of the scale factor due to the vacuum
polarization effect of the quantum fluctuations evolving on that
background geometry.

Throughout the paper we have considered spatially-isotropic
perturbations, \emph{i.e.}, RW geometries and quantum states
compatible with their symmetries (spatial homogeneity and
isotropy). It would be interesting to study the stability of de Sitter
spacetime with respect to general inhomogeneous and anisotropic
perturbations.  In that case one has non-trivial results even for a
massless and conformally-coupled field. The underlying reason is that
when considering inhomogeneous and anisotropic metric perturbations,
the perturbed geometry is no longer conformally flat. Whereas the
stability of Minkowski spacetime with respect to general linear
perturbations has been studied for arbitrary masses and curvature
couplings (see Refs.~\cite{flanagan96,anderson03} and references
therein), such results do not exist for RW backgrounds. In order to
analyze the dynamics of inhomogeneous and anisotropic perturbations
around a spatially-flat RW background one can make use of the
effective action and the semiclassical Einstein equation for general
linear perturbations around a spatially-flat RW background obtained in
Ref.~\cite{campos94} for a massless and conformally coupled field (see
also Ref.~\cite{campos96} for a more compact form). Unfortunately this
linearized semiclassical equation is a complicated
integro-differential equation and its solutions have not been studied
in detail so far%
\footnote{For perturbations around Minkowski it is relatively easy to
  solve the integro-differential equation because it can be
  transformed into a purely algebraic equation by Fourier transforming
  with respect to not only the spatial coordinates but also the time
  coordinate. By contrast, that is not possible in the RW case due to
  the time dependence of the scale factor, and the non-locality in
  time cannot be eliminated.}; see, however, Ref.~\cite{anderson07}
for recent work in this direction.

We have studied the back reaction of the quantum fields on the
dynamics of the spacetime geometry within the framework of
semiclassical gravity, which can be understood as a mean field
approximation where the mean gravitational field is described by a
classical metric whereas its quantum fluctuations are not considered.
In order to study the quantum fluctuations of the gravitational field
one can consider the metric perturbations around a background geometry
corresponding to the semiclassical gravity solution and quantize them
within a low-energy effective field theory approach to quantum gravity
\cite{donoghue94a,donoghue94b,donoghue97,burgess04}. So far this
approach has been mostly applied to weak-field problems
\cite{bjerrum03}, but it seems particularly interesting to
extend its application to strong-field situations involving black
holes and cosmological spacetimes \cite{weinberg05,weinberg06}. The
stochastic gravity formalism \cite{hu03a,hu04a} can be a useful tool
in this respect since one can prove its equivalence to a quantum
treatment of the metric perturbations if graviton loops are neglected,
which can be formally justified in a large $N$ expansion for a large
number of matter fields \cite{hu04b}. A central object in this
formalism is the symmetrized connected two-point function of the
stress tensor operator for the quantum matter fields, which determines
the metric fluctuations induced by the quantum fluctuations of the
matter fields. Such an object has been computed for a massless
minimally coupled field evolving in a de Sitter background spacetime
and the fluctuations of the stress tensor were found to be comparable
to its expectation value \cite{roura99b}. Therefore, studying in
detail the quantum fluctuations of the metric in this context
constitutes a natural extension of our work worth pursuing. The
results obtained here would still be relevant in that case because
they provide the right background around which the metric should be
perturbed and quantized.

We close this section with a brief discussion of the relationship
between our results and the linearization instability for metric
perturbations around de Sitter spacetime coupled to a scalar field
found in Ref.~\cite{losic06}, where it was concluded that it is only
consistent to consider de Sitter invariant states for the quantum
field.  This conclusion does not directly affect our analysis because
we did not consider fluctuations of the metric and studied only the
dynamics of the mean geometry, which couples to the expectation value
of the stress tensor operator of the matter field. The expectation
value of the stress tensor for the class of states that we have
considered in this paper, which are spatially homogenous and
isotropic, automatically satisfies the linearization stability
constraint given by Eq.~(44) in Ref.~\cite{losic06}. It is when
considering the quantum fluctuations of the metric that the
linearization stability condition imposes additional restrictions on
the state of the matter field because in that case the condition must be imposed on the $n$-point correlation
functions of the stress tensor as well,
and this implies that the state of the field must be de Sitter
invariant.%

\begin{acknowledgments}
We are grateful to Daniel Arteaga, Diego Blas, Joan Antoni Cabrer, Dom\`enec
Espriu, Jaume Garriga and Emil Mottola for interesting
discussions. This work has been partly supported by the Research
Projects MEC FPA-2004-04582 and DURSI 2005SGR00082.  A.~R.\ is
supported by LDRD funds from Los Alamos National Laboratory.
\end{acknowledgments}

\appendix*

\section{Semiclassical Einstein equation}
\label{app1}

In this appendix we briefly review the semiclassical Einstein equation
and its derivation by functional techniques.  When neglecting the
effects of graviton loops, the back reaction of quantum matter fields
on the mean gravitational field is described by the semiclassical
Einstein equation, which can be written as
\begin{equation}
G_{ab}\left[ g\right] +\Lambda g_{ab}+ \alpha A_{ab}\left[ g\right]
+\beta B_{ab}\left[ g\right] =\kappa \langle \hat{T}_{ab}\left[
g\right] \rangle_\mathrm{ren}\text{,}
\label{2.8}
\end{equation}
where $G_{ab}$ is the Einstein tensor associated with some globally
hyperbolic spacetime with metric $g_{ab}$, $\langle \hat{T}_{ab}\left[
  g\right] \rangle_\mathrm{ren}$ is the suitably renormalized
expectation value of the stress-tensor operator corresponding to the
scalar field operator $\hat{\phi}\left[ g\right]$, and $\alpha $,
$\beta $, $\Lambda $ and $\kappa$ are renormalized parameters
evaluated at the same renormalization scale as $\langle \hat{T}_{ab}
[g] \rangle_\mathrm{ren}$.  The tensors $A_{ab}$ and $B_{ab}$ are
obtained by functionally differentiating with respect to the metric
terms corresponding to the Lagrangian densities $C^{abcd} C_{abcd}$
and $R^{2}$ in the gravitational action,
\begin{eqnarray}
A^{ab}&=&\frac{1}{\sqrt{-g}}\frac{\delta}{\delta g_{ab}}\int d^4x
\sqrt{-g}\,C_{cdef}C^{cdef},\label{2.9a}\\
B^{ab}&=&\frac{1}{\sqrt{-g}}\frac{\delta}{\delta g_{ab}}\int d^4x
\sqrt{-g}\,R^2,
\label{2.9b}
\end{eqnarray}
where $C_{abcd}$ is the Weyl tensor and $R$ the Ricci curvature
scalar.  These Lagrangian densities $C_{abcd} C^{abcd}$ and $R^{2}$
are related to the counterterms introduced in the bare gravitational
action needed to renormalize the ultraviolet divergencies arising in
the expectation value of the stress tensor.  Note that from their
definitions the tensors (\ref{2.9a})-(\ref{2.9b}) are divergenceless:
$\nabla^a A_{ab}=0=\nabla^a B_{ab}$.

Let us see how the semiclassical Einstein equation can be derived by
functional methods.  To compute the expectation value on the
right-hand side of Eq.~(\ref{2.8}) we can use the closed time path
(CTP) or ``in-in'' functional formalism
\cite{schwinger61,keldysh65,chou85,jordan86,calzetta87,campos94,weinberg05}. Let
us foliate the assumed globally hyperbolic spacetime with
$t=\textrm{const.}$ spacelike hypersurfaces $\Sigma_t$, and denote the
initial and final times by $t_\mathrm{i}$ and $t_\mathrm{f}$,
respectively. In the CTP formalism we introduce two copies of the
metric and the field, $(g_{ab}^+,g_{ab}^-)$ and $(\phi^+,\phi^-)$,
which will coincide at the final time:
$g_{ab}^+(t_\mathrm{f})=g_{ab}^-(t_\mathrm{f})$ and
$\phi^+(t_\mathrm{f})=\phi^-(t_\mathrm{f})$. Let
$\rho_\mathrm{i}[\phi^+(t_\mathrm{i}),\phi^-(t_\mathrm{i}) ]$ be the
matrix element of the density operator describing the initial state of
the scalar field. The Feynman-Vernon influence action
\cite{feynman63,feynman65}, $S_{\textrm{IF}}[g^+,g^-]$, which
describes the effect of the matter field on the gravitational field,
is defined as the following path integral over two copies of the
scalar field:
\begin{equation}
e^{iS_{\textrm{IF}}[g^+,g^-]}=
\int {\cal D}\phi^+{\cal D}\phi^-\rho_\mathrm{i}[\phi^+(t_\mathrm{i}),\phi^-(t_\mathrm{i}) ]
\delta[\phi^+(t_\mathrm{f})-\phi^-(t_\mathrm{f})]
e^{i\left(S_\mathrm{m}[g^+,\phi^+]- S_\mathrm{m}[g^-,\phi^-]\right)},
\label{2.10}
\end{equation}
where $S_\mathrm{m}[g,\phi]$ is the action for the scalar field in the
spacetime described by the metric $g_{ab}$. Neglecting graviton loops,
the CTP effective action for the gravitational field is then
\begin {equation}
\Gamma[g^+,g^-]=S_\mathrm{g}[g^+]-S_\mathrm{g}[g^-]+S_{\mathrm{IF}}[g^+,g^-],
\label{2.11a}
\end{equation}
where $S_\mathrm{g}[g^\pm]$ is the bare gravitational
action. $\Gamma[g^+,g^-]$ is the effective action for the mean
gravitational field coupled to the quantum scalar
field. $S_{\textrm{IF}}[g^+,g^-]$ has ultraviolet divergencies which
can be renormalized by using a suitable regularization procedure and
by adding the aforementioned counterterms to the bare gravitational
action $S_\mathrm{g}[g^\pm]$.  More specifically, one starts with a
regularized gravitational action in Eq.~(\ref{2.11a}) which includes
the bare parameters $\kappa_B, \Lambda_B, \alpha_B, \beta_B$; at the
end of the calculation one takes the regularization parameter to its
physical value and the divergencies are absorbed into the bare
parameters which acquire their dressed physical values. The
renormalized effective action can then be written as
\begin {equation}
\Gamma[g^+,g^-]=S^{\mathrm{ren}}_\mathrm{g}[g^+]-S^{\mathrm{ren}}_\mathrm{g}[g^-]+S^{\mathrm{ren}}_{\mathrm{IF}}[g^+,g^-],
\label{2.11b}
\end{equation}
where the superscript means that these terms have been already
renormalized and are finite.

Since the classical stress tensor of the matter field is defined as
\begin {equation}
T^{ab}= \frac{2}{\sqrt{-g}}\frac{\delta S_\mathrm{m}}{\delta g_{ab}},
\label{2.12}
\end{equation}
one can see from the definition of the influence action in
Eq.~(\ref{2.10}) that the expectation value of the stress tensor in
the given quantum state of the field is given by
\begin {equation}
\langle \hat{T}^{ab}[g]\rangle_{\mathrm{ren}}= \left. \frac{2}{\sqrt{-g}}\frac{\delta S^{\mathrm{ren}}_{\textrm{IF}}[g^+,g^-]}{\delta g_{ab}^+}\right|_{g^+=g^-=g},
\label{2.12a}
\end{equation}
where the renormalized value of the influence action has been used.
Finally, the semiclassical Einstein equations (\ref{2.8}) can be
derived by functional derivation of the effective gravitational
action, $\Gamma[g^+,g^-]$, with respect to $g_{ab}^+$ and then taking
$g_{ab}^+=g_{ab}^-=g_{ab}$:
\begin{equation}
\left. \frac{\delta \Gamma[g^+,g^-]}{\delta g_{ab}^+}\right|_{g^+=g^-=g}=0.
\label{2.14a}
\end{equation}

Notice that doubling the number of fields, the ``plus'' field, which
evolves forward in time, and ``minus'' field, which evolve backwards
in time, is what allows us to obtain an expectation value from the
above functional derivative in the CTP formalism, rather than a
transition element as in the ordinary ``in-out'' effective action
method. This can be clearly seen from the following alternative
representation of the influence action:
\begin{equation}
e^{iS_{\textrm{IF}}[g^+,g^-]}
= \mathrm{Tr} \left[ \hat{U}_{g^+}(t_\mathrm{f},t_\mathrm{i})
\hat{\rho}_\mathrm{i} \hat{U}^\dagger_{g^-}(t_\mathrm{f},t_\mathrm{i}) \right]
=\langle \Psi_\mathrm{i} |\hat{U}_{g^-}(t_\mathrm{i},t_\mathrm{f})
\hat{U}_{g^+}(t_\mathrm{f},t_\mathrm{i})  | \Psi_\mathrm{i} \rangle,
\label{2.13}
\end{equation}
where the last equality holds for a pure initial state of the field
$\hat{\rho}_\mathrm{i} = | \Psi_\mathrm{i} \rangle \langle
\Psi_\mathrm{i} |$ and ${\hat{U}_{g^\pm}}$ is the unitary time
evolution operator for a field $\hat{\phi}$ propagating in a spacetime
with metric $g_{ab}^\pm$:
\begin{equation}
\hat{U}_{g^\pm}(t_\mathrm{f},t_\mathrm{i})=
T\exp\left(-i\int_{t_\mathrm{i}}^{t_\mathrm{f}} dt
\hat{H}_\mathrm{m}[g^\pm,\hat{\phi},\hat{\pi}]\right),
\label{2.14}
\end{equation}
where $T$ denotes time ordering and $\hat{H}_\mathrm{m}[g,\hat{\phi},\hat{\pi}]$
is the Hamiltonian of the scalar field obtained by considering the
spacetime foliation $\{\Sigma_t\}$ and introducing the corresponding
$3+1$ decomposition.

In this paper we use dimensional regularization, this means that the
regularization parameter is $n-4$, where $n$ is the number of
spacetime dimensions. In that case the square of the Weyl tensor in
the Lagrangian density of Eq.~(\ref{2.9a}) must be substituted by
$\frac{2}{3}\left(R_{abcd} R^{\mu\nu\rho\sigma}- R_{ab}R^{ab}\right)$,
where $R_{abcd}$ and $R_{ab}$ are the Riemann and Ricci tensors. Such
a Lagrangian density reduces to the square of the Weyl tensor in $n=4$
when the Gauss-Bonnet theorem is taken into account.

We also consider a massless scalar field throughout.  Using dimensional
regularization the counterterms for a massless scalar field can be
read from the divergent parts in the effective action computed using
the DeWitt-Schwinger expansion of the Feynman propagator in the
coincidence limit; see Eq.~(6.44) in Ref.~\cite{birrell94}. It should
be remarked that the DeWitt-Schwinger expansion is defined for massive
fields and it is ill defined in the massless limit. However, it can
still be used to extract the divergent terms. There are three
divergent terms in this expansion, the first and second terms have
coefficients related to a constant and to the Ricci scalar,
respectively. These two terms vanish in the massless limit. The third
term has a coefficient which is quadratic in the curvature, and in this
case it is convenient to keep the mass as an infrared regulator which
can be removed at the end of the calculation by redefining the
renormalization scale. This term leads to the following counterterm
for the gravitational action:
\begin{eqnarray}
S_\mathrm{g}^\mathrm{c}[g;\mu]&=&
\left(\alpha(\mu) + \frac{\mu^{n-4}}{2880\pi^2(n-4)}\right)\int d^n x\sqrt{-g}\left(R_{abcd}R^{abcd}-R_{ab}R^{ab}\right)\nonumber\\
&&+\left(\beta(\mu) + \frac{\nu^2\mu^{n-4}}{32\pi^2(n-4)}\right)\int d^n x\sqrt{-g}R^2,
\label{2.14b}
\end{eqnarray}
where $\alpha(\mu)$ and $\beta(\mu)$ are the renormalized
dimensionless parameters which appear in Eq.~(\ref{2.8}), and $\mu$ is
a parameter with dimensions of mass which ensures that the action has
the correct dimensions even when $n \neq 4$ and plays the role of the
renormalization scale in dimensional regularization. The bare
parameters $\alpha_\mathrm{B}$ and $\beta_\mathrm{B}$, which
correspond to the whole factor multiplying the first and second
integrals respectively, should be independent of the renormalization
scale $\mu$. Therefore, under a change of the renormalization scale
$\mu \to \mu'$ [and neglecting terms with positive powers of $(n-4)$,
which vanish when $n \to 4$] the renormalized parameters change as
follows: $\alpha(\mu')= \alpha(\mu)-(2880 \pi^2)^{-1} \ln (\mu'/\mu)$
and $\beta(\mu')=\beta(\mu)-(\nu^2 / 32\pi^2) \ln (\mu'/\mu)$.

Note that as seen from Eq.~(\ref{2.29a}), the $R^2$ counterterm in the
action can be easily inferred from the $(n-4)^{-1}$ divergence in
Eq.~(\ref{2.26}), but it is not so obvious how to come up with the
other counterterm (with the Riemann square) because no other
divergence is present. That is because the first term in
Eq.~(\ref{2.25}), which would be the entire contribution for a
conformal field, is finite in the particular case of a conformally
flat metric. However, a slight departure from conformal flatness
(\emph{e.g.}\ by considering inhomogeneous perturbations around the
flat space geometry) renders the first term in Eq.~(\ref{2.25})
divergent and such a divergence is accounted for by the above general
DeWitt-Schwinger expansion \cite{garriga01,garriga03}.  Hence, even
though the first term in Eq.~(\ref{2.25}) would still be independent
of $a$ when regulated with dimensional regularization, the
counterterms that cancel the divergencies in the limit $n \to 4$ would
bring the $a$ dependence into the problem.





\end{document}